\documentclass[tighten,trackchanges]{aastex62}

\usepackage{multirow}
\usepackage{amsmath}

\newcommand{\ud}[2]{\mbox{$^{+ #1}_{- #2}$}}
\graphicspath{{./}{figures/}}
\newcommand{\Chisqred}{\mbox{$\chi^2_\nu$}}
\newcommand{\Chisq}{\mbox{$\chi^2$}}

\newcommand{\ee}[1]{\mbox{$10^{#1}$}}
\newcommand{\tee}[1]{\mbox{$\times 10^{#1}$}}
\newcommand{\percmsq}{\mbox{$\,{\rm cm^{-2}}$}}

\newcommand{\persec}{\mbox{$\,{\rm s^{-1}}$}}

\newcommand{\msun}{\mbox{$\,M_\odot$}}
\newcommand{\cgsflux}{\mbox{$\,{\rm erg\,\percmsq\,\persec}$}}
\newcommand{\cgslum}{\mbox{$\,{\rm erg\,\persec}$}}

\newcommand{\nicer}{\textit{NICER}}

\accepted{for publication in the Astrophysical Journal Letters}

\shorttitle{\textit{NICER} Millisecond Pulsar Data Set}
\shortauthors{Bogdanov et al.}

\begin{document}

\title{CONSTRAINING THE NEUTRON STAR MASS-RADIUS RELATION AND DENSE MATTER EQUATION OF STATE WITH \textit{NICER}. I. THE MILLISECOND PULSAR X-RAY DATA SET}

\author[0000-0002-9870-2742]{Slavko Bogdanov}
\affiliation{Columbia Astrophysics Laboratory, Columbia University, 550 West 120th Street, New York, NY, 10027, USA}

\author[0000-0002-6449-106X]{Sebastien Guillot}
\affil{IRAP, CNRS, 9 avenue du Colonel Roche, BP 44346, F-31028 Toulouse Cedex 4, France}
\affil{Universit\'{e} de Toulouse, CNES, UPS-OMP, F-31028 Toulouse, France}

\author[0000-0002-5297-5278]{Paul S.~Ray} 
\affiliation{Space Science Division, U.S.~Naval Research Laboratory, Washington, DC 20375, USA}

\author[0000-0002-4013-5650]{Michael T.~Wolff}
\affiliation{Space Science Division, U.S.~Naval Research Laboratory, Washington, DC 20375, USA}

\author[0000-0001-8804-8946]{Deepto Chakrabarty}
\affil{MIT Kavli Institute for Astrophysics and Space Research, Massachusetts Institute of Technology, 70 Vassar Street, Cambridge, MA 02139, USA}

\author[0000-0002-6089-6836]{Wynn C.~G.~Ho}
\affil{Department of Physics and Astronomy, Haverford College, 370 Lancaster Avenue, Haverford, PA 19041, USA}
\affil{Mathematical Sciences, Physics and Astronomy, and STAG Research Centre, University of Southampton, Southampton SO17 1BJ, UK}

\author[0000-0002-0893-4073]{Matthew Kerr}
\affiliation{Space Science Division, U.S.~Naval Research Laboratory, Washington, DC 20375, USA}

\author[0000-0002-3862-7402]{Frederick K.~Lamb}
\affil{Center for Theoretical Astrophysics and Department of Physics, University of Illinois at Urbana-Champaign, 1110 West Green Street, Urbana, IL 61801-3080, USA}
\affil{Department of Astronomy, University of Illinois at Urbana-Champaign, 1002 West Green Street, Urbana, IL 61801-3074, USA}

\author{Andrea Lommen}
\affil{Department of Physics and Astronomy, Haverford College, 370 Lancaster Avenue, Haverford, PA 19041, USA}

\author[0000-0002-8961-939X]{Renee M.~Ludlam}\altaffiliation{NASA Einstein Fellow}
\affil{Cahill Center for Astronomy and Astrophysics, California Institute of Technology, Pasadena, CA 91125, USA}

\author{Reilly Milburn}
\affil{Department of Physics and Astronomy, Haverford College, 370 Lancaster Avenue, Haverford, PA 19041, USA}

\author{Sergio Montano}
\affil{Department of Physics and Astronomy, Haverford College, 370 Lancaster Avenue, Haverford, PA 19041, USA}

\author[0000-0002-2666-728X]{M.~Coleman Miller}
\affil{Department of Astronomy and Joint 
Space-Science Institute, University of Maryland, College Park, MD 
20742-2421, USA}

\author{Michi Baub\"ock}
\affil{Max Planck Institut f\"ur Extraterrestrische Physik, Gie{\ss}enbachstr. 1, D-85737 Garching, Germany}

\author{Feryal \"Ozel}
\affil{Steward Observatory, University of Arizona, 933 N.~Cherry Avenue, Tucson, AZ 85721, USA}

\author{Dimitrios Psaltis}
\affil{Steward Observatory, University of Arizona, 933 N.~Cherry Avenue, Tucson, AZ 85721, USA}

\author{Ronald A.~Remillard}
\affil{MIT Kavli Institute for Astrophysics and Space Research, Massachusetts Institute of Technology, 70 Vassar Street, Cambridge, MA 02139, USA}

\author[0000-0001-9313-0493]{Thomas E.~Riley}
\affiliation{Anton Pannekoek Institute for Astronomy, University of Amsterdam, Science Park 904, 1090GE Amsterdam, the Netherlands}

\author{James F.~Steiner}
\affil{Harvard-Smithsonian Center for Astrophysics, 60 Garden Street, Cambridge, MA 02138, USA}

\author[0000-0001-7681-5845]{Tod E.~Strohmayer}
\affiliation{Astrophysics Science Division and Joint Space-Science Institute, NASA Goddard Space Flight Center, Greenbelt, MD 20771, USA}

\author[0000-0002-1009-2354]{Anna L.~Watts}
\affiliation{Anton Pannekoek Institute for Astronomy, University of Amsterdam, Science Park 904, 1090GE Amsterdam, the Netherlands}

\author{Kent~S.~Wood}
\affiliation{Praxis, Arlington, VA, resident at the U.S.~Naval Research Laboratory, Washington, DC 20375, USA}

\author{Jesse Zeldes}
\affil{Department of Physics and Astronomy, Haverford College, 370 Lancaster Avenue, Haverford, PA 19041, USA}

\author[0000-0003-1244-3100]{Teruaki Enoto}
\affil{The Hakubi Center for Advanced Research and Department of Astronomy, Kyoto University, Kyoto 606-8302, Japan}

\author{Takashi Okajima}
\affil{Astrophysics Science Division, NASA Goddard Space Flight Center, Greenbelt, MD 20771, USA}

\author{James W.~Kellogg}
\affil{Applied Engineering and Technology Directorate, NASA Goddard Space Flight Center, Greenbelt, MD 20771, USA}

\author{Charles Baker}
\affil{Applied Engineering and Technology Directorate, NASA Goddard Space Flight Center, Greenbelt, MD 20771, USA}

\author{Craig B.~Markwardt}
\affiliation{X-Ray Astrophysics Laboratory, NASA Goddard Space Flight Center, Greenbelt, MD 20771, USA}

\author{Zaven Arzoumanian}
\affiliation{X-Ray Astrophysics Laboratory, NASA Goddard Space Flight Center, Greenbelt, MD 20771, USA}

\author{Keith C.~Gendreau} 
 \affiliation{X-Ray Astrophysics Laboratory, NASA Goddard Space Flight Center, Greenbelt, MD 20771, USA}

\correspondingauthor{Slavko Bogdanov}
\email{slavko@astro.columbia.edu}

\begin{abstract}
We present the set of deep \textit{Neutron Star Interior Composition Explorer} (\textit{NICER}) X-ray timing observations of the nearby rotation-powered millisecond pulsars PSRs~J0437$-$4715, J0030$+$0451, J1231$-$1411, and J2124$-$3358, selected as targets for constraining the mass-radius relation of neutron stars and the dense matter equation of state via modeling of their pulsed thermal X-ray emission. We describe the instrument, observations, and data processing/reduction procedures, as well as the series of investigations conducted to ensure that the properties of the data sets are suitable for parameter estimation analyses to produce reliable constraints on the neutron star mass-radius relation and the dense matter equation of state. We find that the long-term timing and flux behavior and the Fourier-domain properties of the event data do not exhibit any anomalies that could adversely affect the intended measurements.  From phase-selected spectroscopy, we find that emission from the individual pulse 
peaks is well described by a single-temperature hydrogen atmosphere spectrum, with the exception of PSR~J0437$-$4715, for which multiple temperatures are required.
\end{abstract}


\section{Introduction} \label{sec:intro}

Neutron stars (NSs) provide the only known setting where the regime of ultra-high density, large proton/neutron number asymmetry, and low temperature can be explored. NSs are therefore of tremendous value for nuclear physics, as they offer a path to empirically  determining the state of cold, catalyzed matter beyond nuclear saturation density ($\rho_s= 2.8 \times 10^{14}$\,g\,cm$^{-3}$). Determining the dense matter equation of state (EoS) has far-reaching implications for astrophysics as well. The detailed physics and the accompanying electromagnetic, neutrino, and gravitational wave signals of energetic astrophysical phenomena such as black-hole/NS and double NS mergers, and core-collapse supernovae, are highly sensitive to the interior structure of NSs \citep{Shibata11,Faber12,Read13,delPozzo13,Lackey14,Kumar15,Rosswog15,Bauswein16,Fernandez16,Janka16,Oertel16,Shibata16}.  

Because we cannot directly sample the matter at the core of a NS, we must rely on indirect inference using sensitive observations of their exteriors. Fortunately, the microscopic relation between the pressure $P$ and density $\rho$ of NS matter determines the macroscopic properties of the star, in particular, its radius $R$ and mass $M$ \citep[see for example][]{Lattimer01,Lattimer05,Ozel09,Read09a,hebeler13,OzelFreire16}. This connection between the two relations can, in principle, be exploited via astrophysical observations to derive tight constraints on these parameters in any given parametrization of the EoS \citep[see, e.g.,][and references therein]{2019MNRAS.485.5363G,2019arXiv190408907M}. 
 
This prospect has prompted a number of efforts using a variety of methods to constrain the mass-radius ($M-R$) relation of NSs with X-ray observations,  complementary to those that aim to constrain $R$ and the dense matter EoS using detections of gravitational waves from binary neutron star mergers with the Advanced LIGO and VIRGO gravitational wave observatories \citep{PhysRevLett.121.161101}. In practice, constraining the $M-R$ relation with X-ray observations has proven to be quite difficult due to the absence of strong spectral lines, or ambiguity as to the nature of observed emission features \citep[e.g.,][]{cottam02,changetal06,rauch08,lin10}.  Thermal X-ray radiation from the physical surface of a NS can be used to extract valuable information regarding the EoS \citep[e.g.,][for comprehensive reviews]{heinke13,miller13,ozel13,potekhin14,OzelFreire16}. Although most existing measurements yield values of $R$ generally consistent with the expected range of theoretical values, even for nominally precise measurements there are enough concerns about systematic errors that it is not yet possible to constrain the EoS significantly \citep[see, e.g.,][]{2010ApJ...722...33S,2018MNRAS.476..421S,leahy11,guillot13,2014MNRAS.444..443H,2016EPJA...52...63M,2016ApJ...820...28O,2017A&A...608A..31N}.

For rapidly rotating NSs with the surface X-ray radiation  contained in regions smaller than the whole stellar surface, $R$ and $M$ can be constrained individually through careful modeling of the observed X-ray pulsations. This is possible because the characteristics of the pulsations depend on $R$ and $M$ in different ways \citep{pechenick83,strohmayer92,1995ApJ...442..273P,miller98,braje00,beloborodov02,poutanen03,cadeau07,morsink07,lo13,psaltis14a,psaltis14b,miller15}. 

The \textit{Neutron Star Interior Composition Explorer} (\textit{NICER}; see \citealt{2016SPIE.9905E..1HG}), operating on the \textit{International Space Station (ISS)} since 2017 June, is focusing on measuring $R$ and $M$ of a few nearby rotation-powered MSPs that produce thermal radiation by fitting model pulse profiles to these periodic soft X-ray modulations. These targets have been selected because their X-rays appear to be produced primarily by thermal emission from hotter regions around their magnetic poles. The pulsations are always present, the beaming pattern and spectrum of the emission that produces them is thought to be relatively well understood, and the rotation rates of these stars are rapid and exceptionally stable. Moreover, based on simulations, it is expected that  (unlike the case for other methods) if a fit to the joint phase and energy properties of \textit{NICER} MSP X-ray pulsations is statistically good, it is not strongly biased \citep[see, e.g.,][]{lo13,miller15}. Rotation-powered MSPs are, in this regard, more favorable for $R$ and $M$ measurements using \textit{NICER} than are the modulations produced by {\em i)} accretion-powered millisecond X-ray pulsars, which exhibit temporally varying pulsation properties with no widely accepted model of their X-ray emission, including non-thermal processes from poorly understood regions on and above the stellar surface (see, e.g., \citealt{hartman08,patruno12}); or {\em ii)} burst oscillation sources, which suffer, by comparison, from being extremely transient and from an uncertainty as to whether the hot spots always ignite in the same location (see, e.g., \citealt{watts12} for a review).

The present article is the first in a series of papers describing the data, model, and methodology for obtaining constraints on the NS $M-R$ relation and the dense matter EoS. Here, we describe the targeted millisecond pulsars, their observations, and data sets obtained with \nicer{}, the analyses of which will be published in subsequent works. In \cite{bogdanov19b}, Paper II, we present the approach and codes we use to describe the propagation of the photons emitted from the surface to the observer, while \cite{bogdanov19c}, Paper III, describes all other aspects of the modeling technique applied to the \textit{NICER} data and the potential sources of systematic error. The first set of results, for PSR J0030$+$0451, of the parameter estimation analyses that are based on the data described here to obtain estimates on $M$ and $R$, as well as the dense matter equation of state are presented in \cite{miller19}, \cite{riley19}, and \cite{raaijmakers19}. Results for the other targets will be presented in subsequent publications.  The work is organized as follows: in Section~\ref{sec:nicer}, we describe the \textit{NICER} telescope and its performance. In Section~\ref{sec:reduction}, we detail the observations and data reduction procedures used to obtain the MSP event lists used for parameter estimation analyses. Section~\ref{sec:background} deals with non-source background emission specific to \textit{NICER} and the methods used to estimate it. In Section~\ref{sec:msps} we provide a brief overview of the history of X-ray observations of rotation-powered MSPs. In Section~\ref{sec:targets}, we discuss the four targeted MSPs and the corresponding \textit{NICER} spin phase-folded data. Section~\ref{sec:timing} focuses on the event folding, long-term timing, and Fourier-domain properties of the event data. In Section~\ref{sec:spectroscopy}, we present phase-selected spectroscopy of the four MSPs. We offer conclusions in Section~\ref{sec:conclusions}.

\section{The \textit{NICER} XTI Instrument Performance and Calibration}
\label{sec:nicer}

The \textit{NICER} X-ray Timing Instrument (XTI) consists of an array of 52 active silicon drift detectors housed in focal plane modules (FPMs), each paired with a nested single-reflection grazing-incidence ``concentrator'' optic assembly in the optical path. Groups of eight FPMs are controlled by a single Measurement and Power Unit (MPU). The XTI's concentrator optics are co-aligned, collecting sky emission from a single $\approx$3$'$ radius non-imaging field of view. The instrument is sensitive to X-rays in the 0.2--12\,keV band, with a peak effective area of $\approx$1900\,cm$^{2}$ around 1.5\,keV \citep{2016SPIE.9905E..1HG}. The lower bound is dictated by 
absorption in optical-blocking filters, electronic noise in the cooled detectors, and increasing optical light-loading noise at the lowest energies, while the upper bound is driven by a decline in grazing-incidence reflectivity as well as the quantum efficiency of the silicon detectors.

Photons or charged particles incident on the XTI silicon drift detectors induce an amplified charge signal, which is processed in parallel by a slow and fast analog chain with 465\,ns and 85\,ns shaping time constants, respectively \citep{2016SPIE.9905E..1IP}. The slow chain provides a more precise energy measurement, while the fast chain provides a more precise arrival time measurement of the incident event. Signals that are above a preset threshold in each chain produce an electronic trigger that causes the arrival time and pulse height amplitude of the incident event to be sampled and digitized.  Events that cause both chains to trigger have their fast chain timestamp reported; otherwise, the slow chain timestamp is reported. Whether the chain triggers depends on the event pulse height (which is approximately proportional to the energy deposition within the detector), such that events with energies $E\lesssim 1$\,keV do not trigger the fast chain, while higher-energy events will trigger both chains. This includes X-rays as well as energetic particles and $\gamma$-rays produced by particle interactions with the detector or surrounding structure of \textit{NICER}.

The fast chain timing uncertainty is 70\,ns. The slow minus fast timing uncertainty is $< 4$\,ns, so that the two analog chains have nearly identical timing uncertainties. Time biases are typically $\sim$250\,ns for the fast chain and $\sim$760\,ns for the slow chain for individual \textit{NICER} detectors, and timing variations between individual detectors are typically $\sim$11\,ns. The measured biases are corrected using \textit{NICER} standard software ({\tt nicertimecal}). The \textit{NICER} calibrated event timestamp values after this calibration is performed refer to the time that an on-axis X-ray or particle entered the detector aperture. The time stamp of an event is referenced to the GPS receiver on \textit{NICER}. For the intended analysis of the MSP pulse profiles, the time binning is the pulse period of a few ms divided by 32, compared to which any \textit{NICER} time-tagging uncertainties are negligible.

For the analyses of the \textit{NICER} data presented here, we used products from the calibration database (CALDB) version 20181105 and gain solution (the relationship between energy deposition and pulse height) version {\tt optmv7}. The on-axis effective area including all 52 active detectors is shown in Figure~\ref{fig:nicer_response}. As described in Appendix~\ref{app:optimize}, for PSRs~J0437$-$4715 and J2124$-$3358 we use offset pointings to minimize contamination from neighboring background sources. Because these pulsars are observed off axis, for their parameter estimation analyses it is necessary to consider an effective area curve that accounts for the resulting decline in sensitivity. For this purpose, we conducted observations of the Crab Nebula and pulsar both on axis and at an offset matching that used for the PSR~J0437$-$4715 observations. On- and off-axis effective area curves derived from ray tracing simulations, and the ratio between off- and on-axis Crab spectrum measurements, are shown in Figure~\ref{fig:nicer_response}. 

Calibration of the effective area of the \textit{NICER} XTI was carried out using observations of the Crab. The energy-dependent residuals in the fits to the Crab spectrum are typically at the level of $\lesssim2\%$, likely stemming from lack of knowledge of the detailed microphysics of the concentrator optics. Efforts are under way by the \textit{NICER} calibration team to further improve the instrument model.

%
%
\begin{figure}
\centering
\includegraphics[clip, angle=0,width=0.48\textwidth]{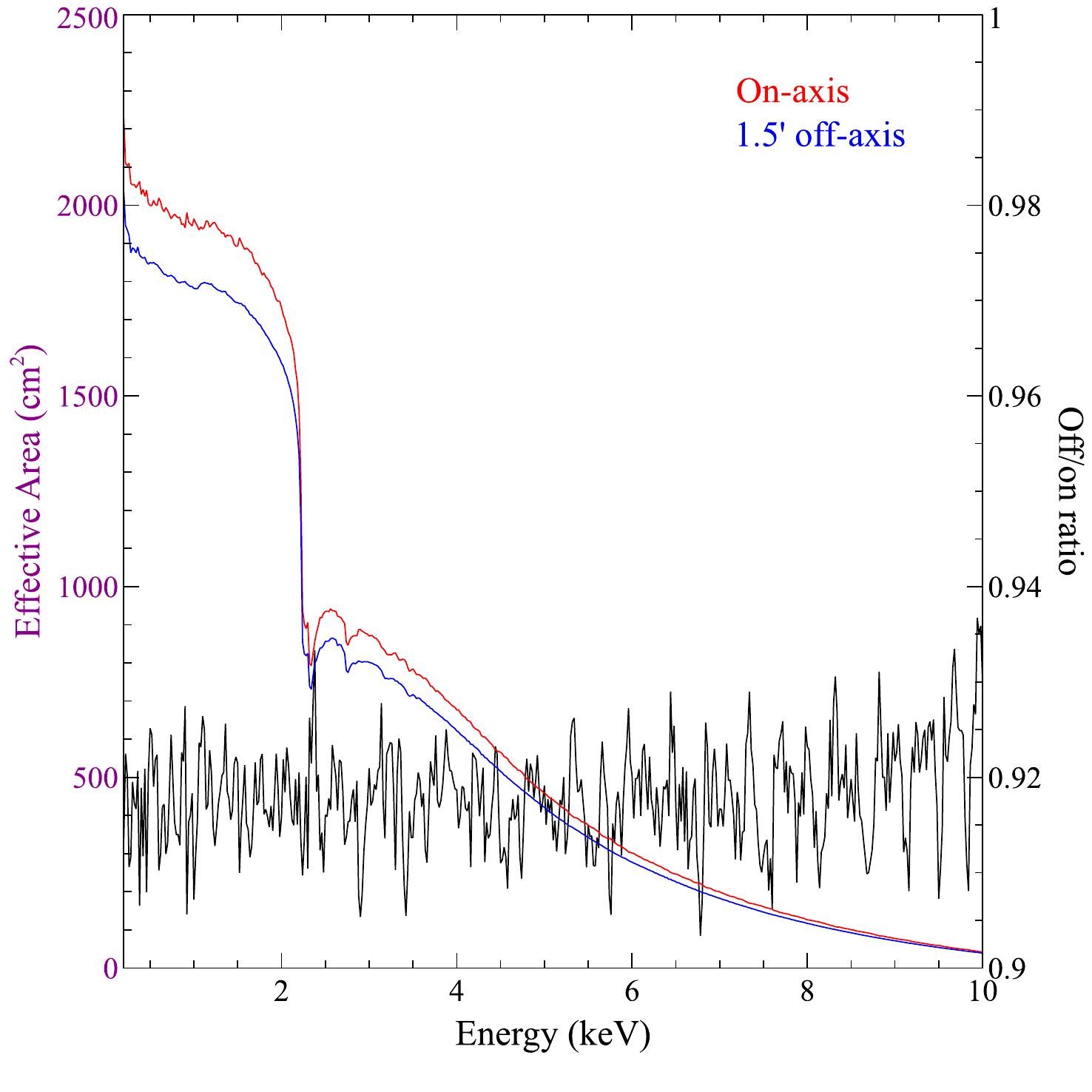}~~~~
\caption{Effective area (red line) of the \textit{NICER} XTI concentrator optics as a function of photon energy, derived from ray-tracing simulations and neglecting low-energy absorption from filters in the optical path; also shown is the estimated effective area (blue line) at $1\farcm 5$ off axis, appropriate for the \textit{NICER} observations of PSR~J0437$-$4715 (see Appendix~\ref{app:optimize} for details). The reduction in sensitivity of the off-axis response has weak or no energy dependence, as shown by the ratio of the off- and on-axis spectrum ratio (black line) measured via observations of the Crab Nebula and pulsar; variations from channel to channel are due to statistical fluctuations. }
\label{fig:nicer_response}
\end{figure}

\section{Observations and Data Reduction}\label{sec:reduction}

The data sets considered here were acquired over the period starting in 2017 June through 2019 June. Owing to the 92-minute orbit of the \textit{ISS}, events are typically accumulated in a large number of separate exposures, each lasting several hundred to $\sim$2000 s. Exposures obtained during the same UTC days are grouped into a single observation (ObsID). The observations for the four MSPs discussed here are summarized in Table~\ref{tab:obsid}.  The data processing and filtering was performed using HEASoft 6.25\footnote{\url{https://heasarc.nasa.gov/lheasoft/}} and NICERDAS version 5.0.  For all sources, the initial event lists are subjected to the same standard filtering criteria:


\begin{deluxetable}{lrrc}
\tablecolumns{4} 
\tablecaption{\textit{NICER} observations of the four millisecond pulsars studied here \label{tab:obsid}}
\tablehead{
\colhead{ } & \colhead{ObsID} & \colhead{Raw}  & \colhead{Total filtered} \\
\colhead{Pulsar} & \colhead{range} & \colhead{exposure (Ms)}  & \colhead{exposure (Ms)}
}
\startdata
PSR~J0437--4715 &   0060010101 -- 0060010110 & 0.071 & \multirow{3}{*}{$0.951$} \\
				& 	1060010101 -- 1060010439 & 2.098 & \\
				& 	2060010401 -- 2060010405 & 0.032 & \\
\hline
PSR~J0030+0451  &   1060020101 -- 1060020437 & 3.076  & $1.936$ \\
\hline
PSR~J1231--1411 &   0060060101 -- 0060060113 & 0.108 & \multirow{3}{*}{1.356} \\
				&   1060060101 -- 1060060373 & 1.982
 & \\
				&   2060060301 -- 1060060389 & 0.395 & \\
\hline
PSR~J2124--3358 &   0060040101 -- 0060040104 & 0.003 & \multirow{3}{*}{1.051} \\
				&   1060040101 -- 1060040313 &  1.377 & \\
				&   2060040301 -- 1060040348 & 0.288 & \\
\enddata
\tablecomments{The exposure time columns report the total duration of data collection (``Raw''), and the exposure time after the filtering described in Section~\ref{sec:reduction}.} 
\end{deluxetable}

\begin{itemize}

\item \textbf{Maximum angular distance from target} -- For standard \textit{NICER} science analyses, event data for a particular target are considered valid if the XTI boresight is within 0.015$^{\circ}$ of the source position. This same criterion is applied to the MSP data we consider here.

\item \textbf{South Atlantic Anomaly passages and particle background excision} -- Particle background is particularly severe during times when the \textit{ISS} is near the South Atlantic Anomaly (SAA). Particle-induced events typically have very high amplitudes and/or occur far from the center of the detector. Many of these events can be filtered out of the event list to produce a cleaned list of predominantly X-ray--only events using the detected amplitude and/or offset from the detector center. This is possible because the entrance aperture of each FPM for X-rays is only 2\,mm in diameter, while the entire active area of the  physical detector is 25\,mm$^2$ \citep{2016SPIE.9905E..1IP}. The ratio of pulse invariant (PI) amplitudes for events detected in both the slow and fast chains, PI\_RATIO $=$ PI\_SLOW$/$PI\_FAST, is related to the event offset from the center of the detector. Events with PI\_RATIO $>$ 1.1 $+$ 120$/$PI are likely particle events and are normally excluded from standard X-ray analysis. This is called ``trumpet'' filtering because the PI versus PI\_RATIO cloud resembles a trumpet\footnote{See, e.g., Figure~6 at \url{https://heasarc.gsfc.nasa.gov/docs/nicer/mission_guide/}.}.

\item \textbf{Minimum elevation above Earth limb and bright Earth limb} -- Observations with a pointing direction close to the limb of the Earth can be strongly affected by optical light-loading noise, especially during orbit day (bright Earth). Therefore, events are filtered to only include intervals when the elevation of the pointing direction above the Earth's limb is $>20^{\circ}$ and $>30^{\circ}$ above the bright Earth  (ELV $>20^{\circ}$ and BR\_EARTH $>30^{\circ}$).

\item \textbf{Exclusion of times of bad tracking} -- \textit{NICER} event data for a particular source are included only if the telescope is on-target. The required conditions for good source tracking are determined using parameters provided in the auxiliary ``make filter'' (MKF) file, specifically  ATT\_MODE=1, ATT\_SUBMODE\_AZ=2, ATT\_SUBMODE\_EL=2. In addition, times when the star tracker solution is not valid are filtered out, when the condition ST\_VALID=1 is not met.

\end{itemize}

The data sets were further screened using additional event filtering criteria, tailored for the purposes of pulse profile modeling. These include:
\begin{itemize}

\item \textbf{Minimum number of enabled detectors} -- For a variety of reasons, not all FPMs are actively registering events at all times. In the standard \textit{NICER} data processing, the default filtering removes all good time intervals (GTIs) with fewer than 38 active detectors (defined through the MIN\_FPM parameter).  However, to ensure uniformity of the MSP data sets such that the effective area over time is constant (thus only requiring a single, time-average effective area curve in our analyses), we imposed a stricter requirement of accepting only GTIs during which all 52 FPMs are active.

\end{itemize}

For the intended analysis of these data, we are interested in the thermal X-rays that originate from the surface of the observed NSs. However, a significant subset of the events collected for each target have different origins and contribute to a background emission component. Both particles and local (non-cosmic) high-energy photons can generate events detected by the FPMs.  There are a number of sources of local electromagnetic background radiation that can affect \textit{NICER} X-ray data. Some of this background is removed by the standard processing pipeline, but residual background events may remain in the cleaned data.  For instance, since \textit{NICER} XTI is a non-imaging instrument, a portion of the emission comes from the unresolved diffuse X-ray background as well as other point sources that fall within the $\sim$6\arcmin\ diameter telescope field of view (FOV); see Figures~\ref{fig:vignetting} and \ref{fig:images}. In Appendix~\ref{app:optimize} we examine this contribution to the background for each MSP, in order to determine the optimal pointing that provides the best signal-to-noise ratio (S/N). Solar wind charge-exchange also contributes to the background and occurs when a high charge state ion in the solar-wind exchanges charge with a neutral species; the resultant ion is in an excited state and in the transition to the ground state can emit a soft X-ray photon.

There are other sources of background emission that are specific to \textit{NICER} and which motivated additional filtering:

\begin{itemize}

\item \textbf{Exclusion of ``hot'' detectors} -- Three of the active FPMs (DET\_ID 14, 34, and 52) are often found to exhibit count rates well above the average of the other detectors, with DET\_ID 34 exhibiting such behavior most frequently. For this reason, from the GTIs with 52 active FPMs we removed events from these ``hot'' detectors. Removing only events from detector 34 produces similar clean exposure times and count rates compared to removing all three. In \cite{miller19}, the \textit{NICER} data  for PSR~J0030+0451 were cleaned by removing DET\_ID 34, while in \cite{riley19}, the data with DET\_ID 14, 34, and 52 removed were analyzed. The effective area was reduced by the appropriate amount ($51/52$ or $49/52$) to account for the use of fewer detectors.

\item \textbf{Sun angle limits} -- Sunlight produces detected noise events at low offset angles from the Sun, and potentially also for low bright-Earth offset angles. In addition, sunlight reflecting off of the ISS solar panels or other ISS structures can be reflected into a subset of \textit{NICER} FPMs. Solar radiation, whether direct or reflected, usually affects the low energy ($E \ll 1$~keV) portion of the spectrum. The result is a light-loading noise peak at very low ($E < 0.3$~keV) energies. This noise peak exhibits a Gaussian-like energy distribution with a variable amplitude. When the amplitude is large, the tail of the Gaussian distribution may leak into higher energies, up to $\sim$0.4$-$0.5~keV. This is common in certain, especially light-sensitive detectors, notably FPM 34.  This additional low energy  background is not desirable since the MSP targets considered here have relatively soft spectra. Based on this, we filter the data such that we only include observations obtained at angles greater than $80^{\circ}$ with respect to the Sun (SUN\_ANGLE$>$80).

\item \textbf{Variable background filtering} -- The target MSPs considered here are expected to show no short- or long-term flux variability in their surface thermal emission. Therefore, any variability observed in the \textit{NICER} data has a non-source origin and is usually due to local (non-cosmic) radiation and particle events.  Even after standard filtering, short-lived instances (lasting seconds to minutes) of intense background flaring reaching count rates up to $\sim$100 count\,s$^{-1}$ are occasionally still present in the data. Such intervals are excised from the event lists by constructing a binned time series light curve with a 16\,s resolution in the 0.25--8~keV band and applying a filter that removes all time bins that exceed a threshold count rate. The count rate cuts applied for PSRs~J0437$-$4715,  J0030+0451, J1231$-$1411, J2124$-$3358 are 3.5, 3.0, 3.0, and 2.8\,count\,s$^{-1}$, respectively, which correspond approximately to a cutoff at $+2\sigma$ from the mean rate.

\item \textbf{Filtering by photon energy} -- For all targets, we limit our analysis to events above 0.25~keV (corresponding to detector channel $\ge 25$), since at lower energies there is increased noise from optical loading and there is greater uncertainty in the triggering efficiency for events. Because the MSPs under consideration have relatively soft spectra, for the parameter estimation analyses we also ignore all events above 3~keV (detector channel $\ge 300$), where the thermal emission becomes negligible and the non-source background greatly dominates.  

\end{itemize}

\section{Background Modeling}
\label{sec:background}

As noted above, a portion of the non-source background contained in a typical \textit{NICER} data set originates from the local environment of the telescope. The \textit{NICER} team has developed two distinct approaches for modeling this time-dependent background emission. 

The first method relies on a combination of two indicators of the space-weather environment, which are found to correlate closely with observed  \textit{NICER} background levels. 

\begin{itemize}
\item \textbf{The Cut-Off Rigidity}, as originally defined for the \textit{BeppoSAX} mission (COR\_SAX; see \citealt{2014ExA....37..599C} and references therein), is a measure of the minimum momentum per unit charge (expressed in units of GeV$/$c) a particle must have in order to reach a certain geographical location. Therefore, as defined, a lower COR\_SAX value is an indication of higher background due to an increased influx of lower momentum particles. 

\item \textbf{The planetary $K$-index ($K_p$)} is commonly used as a measure of geomagnetic storm activity and aurora strength; it quantifies disturbances in the horizontal component of Earth's magnetic field and is expressed as an integer in the range $0-9$, with higher numbers indicating more activity \citep{LINCOLN196767}.

\end{itemize}

This ``environmental'' background model also uses the SUN\_ANGLE parameter, which helps describe the low-energy background produced by optical loading.  COR\_SAX and SUN\_ANGLE are contained in the MKF file (either the standard auxil/ni*.mkf file distributed with processed data or the augmented MKF file produced by the {\tt niprefilter2} tool distributed with the NICERDAS HEASoft package). This background creation method uses two files:
    {\em i)} the background events file\footnote{Current version: \url{https://heasarc.gsfc.nasa.gov/FTP/caldb/data/nicer/xti/pcf/30nov18targskc_enhanced.evt}} 
    that serves as a reference library, and 
    {\em ii)} the KP.fits file\footnote{The most current version, updated daily, is available at \url{https://heasarc.gsfc.nasa.gov/FTP/caldb/data/gen/pcf/kp.fits}}.

The diffuse cosmic X-ray background in the \textit{NICER} blank fields 
(pointings of the {\it Rossi X-Ray Timing Explorer} background fields, \citealt{2006ApJS..163..401J})
is included in the estimated model background as an average of exposures of seven blank fields, with weighting of the averaging adjusted to match the $K_p$ and COR\_SAX of the target data.
The contribution from point sources within the specific FOV near the target is not included, hence if the latter is important it must be treated case by case, by dealing with the relevant field sources. To produce a background estimate for a particular observation, the background reference library is used to find data from prior observations of blank fields with similar combinations of COR\_SAX, $K_p$, and SUN\_ANGLE values, and interpolating between the tabulated values. This modeling approach is predictive, in the sense that it does not rely on any of the event data of the source under consideration.

The second background measurement technique (referred to as the ``3C50'' model) uses the actual source event data to estimate in-band background by matching background library entries with observed event rates in {\em i)} the 15--17~keV range, where the performance of the XTI is such that effectively no astrophysical signal is expected, {\em ii)} a region in PI--PI\_RATIO\footnote{The PI\_RATIO  is defined as the ratio of the PI values measured by slow and fast chains, PI\_RATIO $=$ PI\_SLOW$/$PI\_FAST.} space selected to capture the non-focused background, and {\em iii)} the slow chain noise band ($<0.2$~keV).  The matching is done on 120\,s intervals, and then an exposure-weighted sum of library spectra is computed. This background method has been implemented in HEASoft through the \texttt{nibackgen3C50} command.

For the spectroscopic analyses in Section~\ref{sec:spectroscopy}, we employ the space weather-based background models. We note that in the detailed parameter estimation analyses for PSR~J0030$+$0451 presented in \cite{miller19} and \cite{riley19}, no estimated background is explicitly taken into account; instead, non-hot-spot emission in each detector channel is treated as a free parameter and is assumed to have no dependence on spin phase. The space weather-based background estimate is used as a lower bound
on the total emission that does not originate from the hot regions on the NS surface.

\section{Rotation-Powered Millisecond Pulsars}
\label{sec:msps}

Rotation-powered (``recycled'') MSPs are a population of old NSs ($\sim 10^9$ yr), characterized by rapid rotation rates (a few hundred Hz), exceptional rotational stability, and low inferred dipole magnetic fields ($\sim$10$^{8-9}$ G). These NSs are commonly believed to arise from slowly rotating pulsars in low-mass X-ray binaries \citep{alpar82,radhakrishnansrinivasan82}, which acquire rapid spin rates via accretion of matter and angular momentum. At the end of their spin-up phase they are reactivated as rotation-powered (radio and $\gamma$-ray loud) pulsars, meaning that the observed radiation is generated at the expense of the rotational kinetic energy of the NS. Rotation-powered MSPs were identified as pulsed X-ray sources by  \citet{becker93} in observations with \textit{ROSAT}. 

Over the past two decades, extensive studies with \textit{Chandra} and \textit{XMM-Newton} have shown that many of these NSs are detected as X-ray sources due to thermal emission with temperatures of $\sim10^6$\,K \citep{zavlin06,bogdanov06,bogdanov11,forestell14}. The inferred emitting areas indicate that this radiation is localized in regions on the stellar surface that are much smaller than the total surface area, but comparable to what is expected for pulsar magnetic polar caps. This finding is consistent with pulsar electrodynamics models, which predict heating of the polar caps by a backflow of energetic particles along the open magnetic field lines \citep{harding02,Lockhart2019}. The potential utility of recycled MSPs as powerful probes of the NS structure was first pointed out by \cite{pavlov97} and \cite{zavlin98}, who used \textit{ROSAT} data of the nearest known MSP, PSR~J0437--4715 \citep{johnston93}, to demonstrate that a model of polar cap thermal emission from a NS hydrogen atmosphere provides an adequate description of the X-ray pulse profiles of this MSP, as well as to place crude limits on the $M$-$R$ relation.

Prompted by this promising result, deep \textit{XMM-Newton} timing observations of nearby MSPs were conducted, which confirmed that a non-magnetic hydrogen atmosphere can reproduce the energy-dependent X-ray pulse profiles of the two closest known MSPs, PSRs~J0437$-$4715 and J0030$+$0451.  In contrast, the large-amplitude pulsations were found to be incompatible with a model that considers an isotropically-emitting Planck spectrum. Furthermore, this modeling has already produced some constraints on the allowed NS $M$-$R$ relation.  For PSR~J0437$-$4715, assuming $1.44$\msun (the current best measurement from radio timing, including Shapiro delay measurements; \citealt{reardon16}) the stellar radius is constrained to be $R>10.7$ km (at 3$\sigma$ confidence; \citealt{bogdanov13}), while for the isolated PSR~J0030+0451 the best constraint is $R > 10.4$ km (at 99.9\% confidence) assuming $1.4$\msun  \citep{bogdanov09}. Although these existing limits are not particularly stringent they have nevertheless served to demonstrate the feasibility of this approach and have motivated the deep \textit{NICER} observations described here.
  
\begin{deluxetable}{lclrcccccc}
\tablecolumns{9} 
\tablecaption{Millisecond Pulsars Selected for $M-R$ and EoS Constraints Using \textit{NICER}
\label{tab:MSPlist}}  
\tablehead{\colhead{PSR} & \colhead{$P$} & \colhead{$\dot{P}$\tablenotemark{a}} &     \colhead{$D$\tablenotemark{b}} &    \colhead{$P_b$} & \colhead{$M_{\rm NS}$} & \colhead{$M_{c}$} & \colhead{$F_X$\tablenotemark{c}} & \colhead{\textit{NICER} rate\tablenotemark{d}} &  \colhead{Refs.} \\
           \colhead{ }      & \colhead{(ms)} & \colhead{($\times 10^{-20}$)} & \colhead{(pc)} & \colhead{(d)} & \colhead{(\msun)} & \colhead{(\msun)} & \colhead{(erg cm$^{-2}$ s$^{-1}$)} & \colhead{(ks$^{-1}$)} &  \colhead{}   
} 
\startdata
J0437$-$4715 &	5.76	&	$1.37$	&	$156.79(25)$	&	$5.741$	&	$1.44(7)$	& $0.224(7)$ & 	$1.29\times10^{-12}$	&	1430 & 1,2	\\
J0030$+$0451	    &	4.87	&	$1.02$	&	$325(9)$	&	\nodata	&	\nodata & \nodata	        &	$2.8\times10^{-13}$	&	$314$   & 3,4 \\
J1231$-$1411	&	3.68	&	$0.76$	&	$420$	&	$1.860$	&		unknown    & $\ge$0.19    &	$1.2\times10^{-13}$	&	$210$    & 5 \\
J2124$-$3358	&	4.93	&	$0.73$	&	$410^{+90}_{-70}$	&	\nodata	&	\nodata	 & \nodata      &	$1.7\times10^{-13}$	&	$110$  &   6,7,2 \\
\enddata
\tablerefs{(1) \citealt{johnston93} (2) \citealt{reardon16} (3) \citealt{lommen00} (4) \citealt{2018ApJS..235...37A} (5) \citealt{2011ApJ...727L..16R} (6) \citealt{bailes97} (7) \citealt{lynch2018} 
}
\tablenotetext{{\rm a}}{Intrinsic spin-down rates, corrected for proper motion.}
\tablenotetext{{\rm b}}{Distances with quoted uncertainties are based on parallax measurements. For PSR~J1231$-$1411, the distance is estimated from its dispersion measure and the \citet{YMW16} electron density model of the Galaxy.}
\tablenotetext{{\rm c}}{Unabsorbed source energy flux in the 0.25--2~keV band.}
\tablenotetext{{\rm d}}{\textit{NICER} source count rate per ks in the 0.25--10~keV band.}
\end{deluxetable}

\section{The \textit{NICER} Millisecond Pulsar Target Sample} 
\label{sec:targets}

We now shift focus to the four brightest MSPs selected as primary targets for $M$-$R$ constraints with \textit{NICER}.  For each pulsar we provide a brief overview of the relevant characteristics that make it an important \textit{NICER} target and prior X-ray observations, and present the data obtained thus far. The spin parameters and binary properties (orbital period, NS mass, and companion mass, where applicable) of these pulsars are summarized in Tables~\ref{tab:MSPlist}. In \cite{guillot19}, we present \textit{NICER} observations of other nearby rotation-powered MSPs conducted to assess their potential for providing additional $M$-$R$ constraints in the future. In \cite{arzoumanian19}, we present the \textit{NICER} detection of thermal X-ray pulsations from PSRs~J1614$-$2230 and J0740$+$6620, two of the three most massive NSs known ($M\approx 2$ M$_{\odot}$). These targets will be the subject of parameter estimation analyses for $M-R$ and dense matter equation of state constraints in subsequent publications.

%
%
\begin{figure}
\centering
\includegraphics[clip,trim = 3cm 5cm 1cm 3cm, angle=0,width=0.42\textwidth]{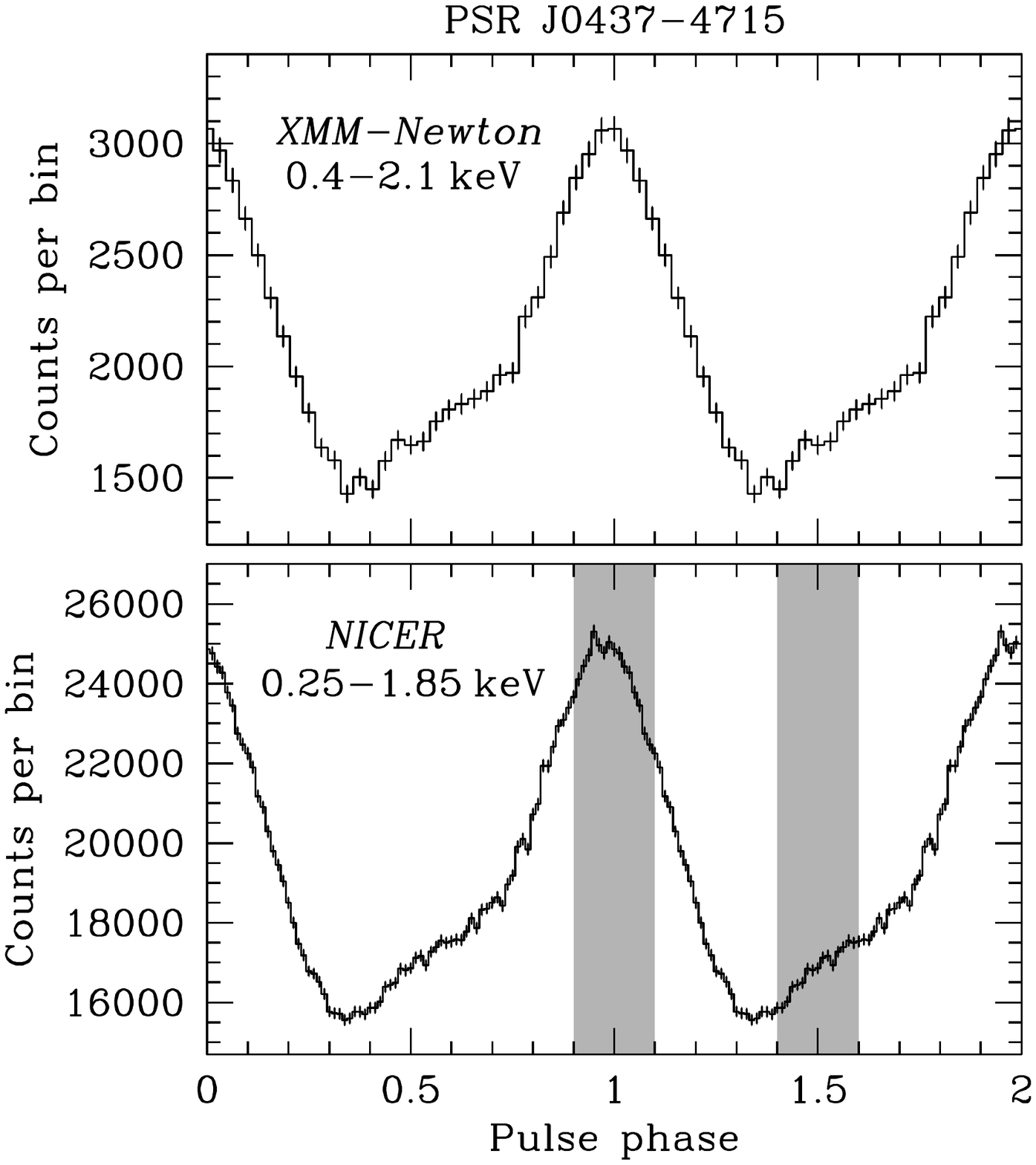}~~~~
\includegraphics[clip,trim = 3cm 5cm 1cm 3cm, angle=0,width=0.42\textwidth]{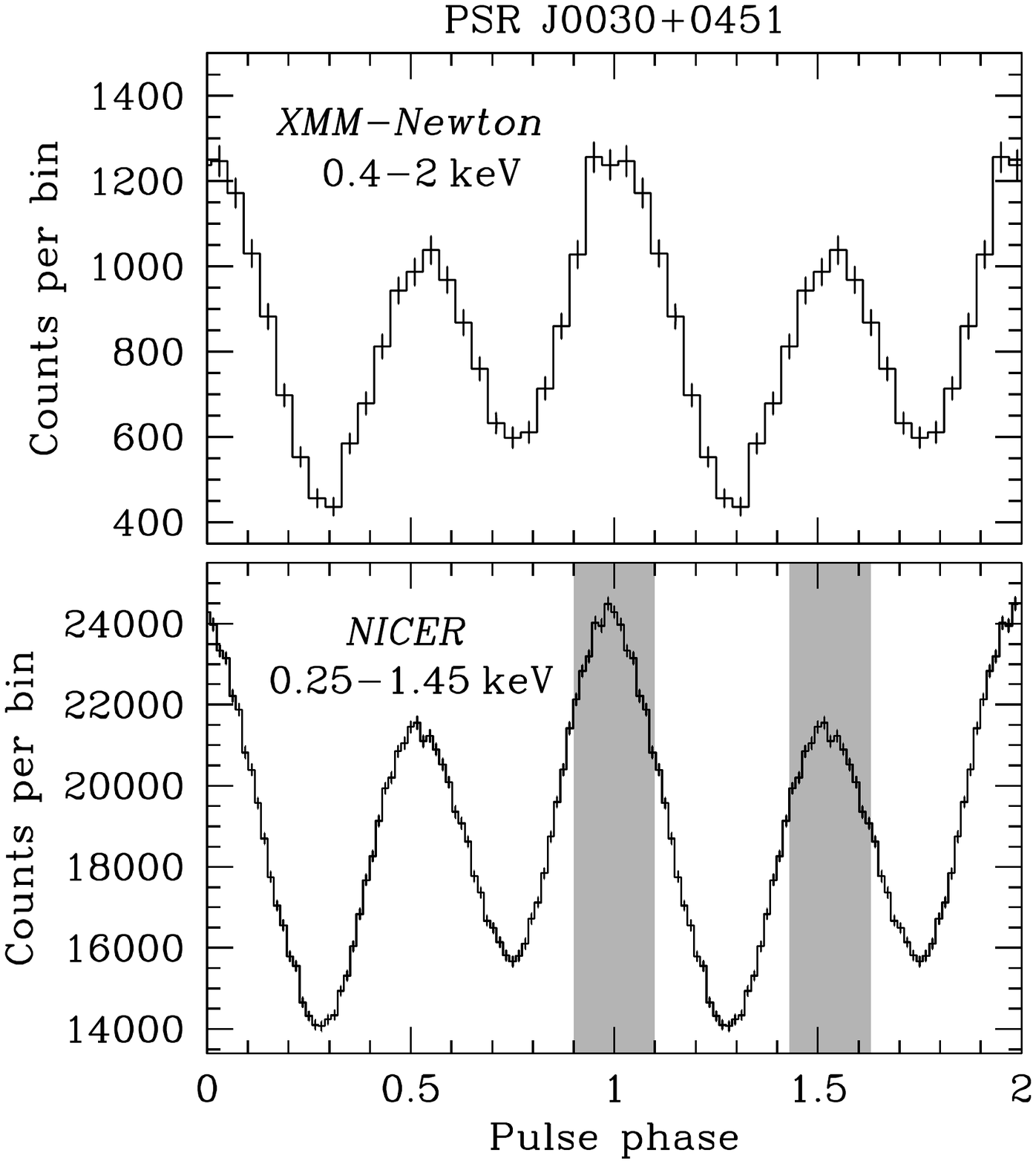}
\includegraphics[clip,trim = 3cm 13.5cm 1cm 3cm, angle=0,width=0.42\textwidth]{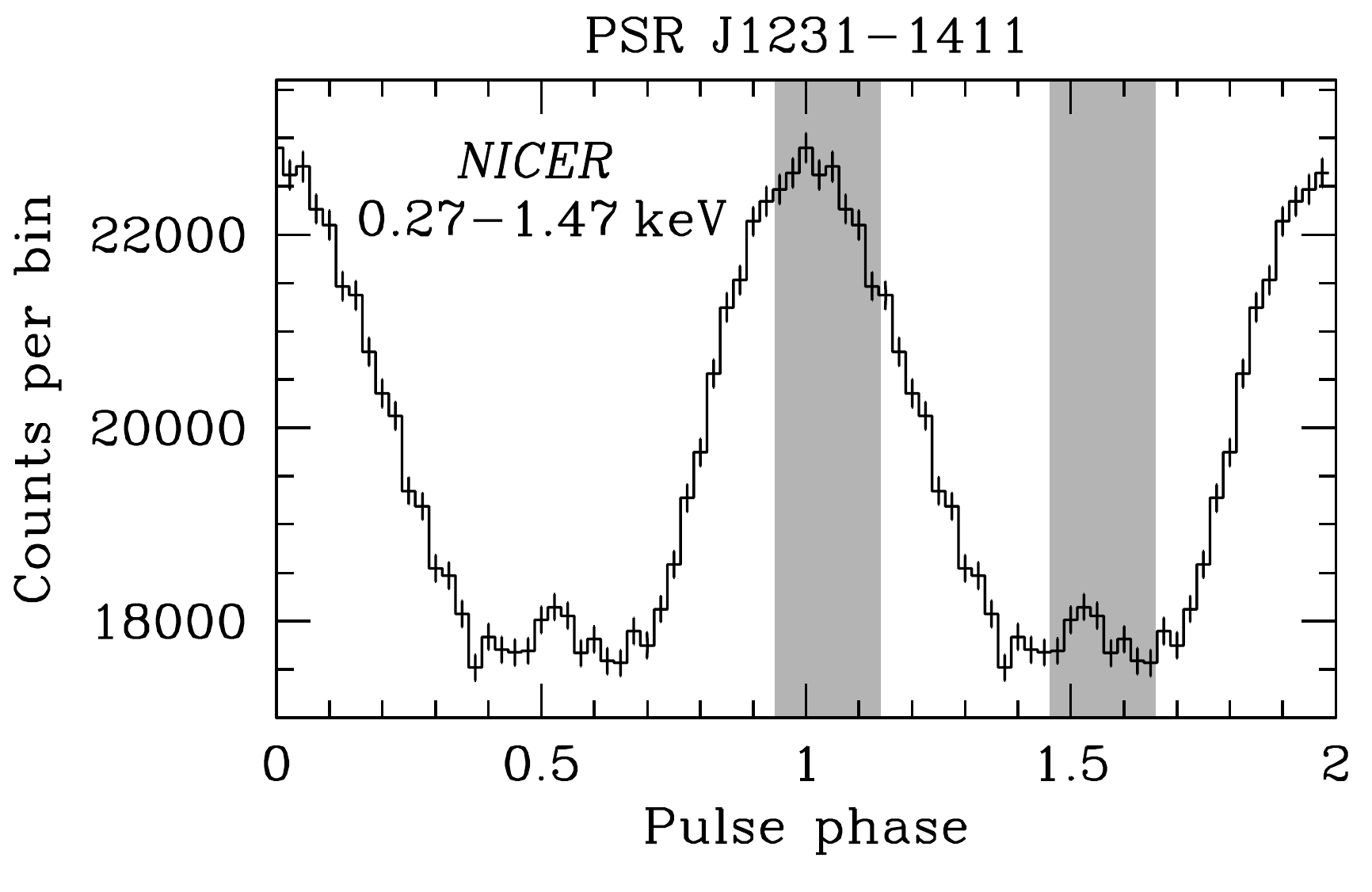}~~~~
\includegraphics[clip,trim = 3cm 5cm 1cm 3cm, angle=0,width=0.42\textwidth]{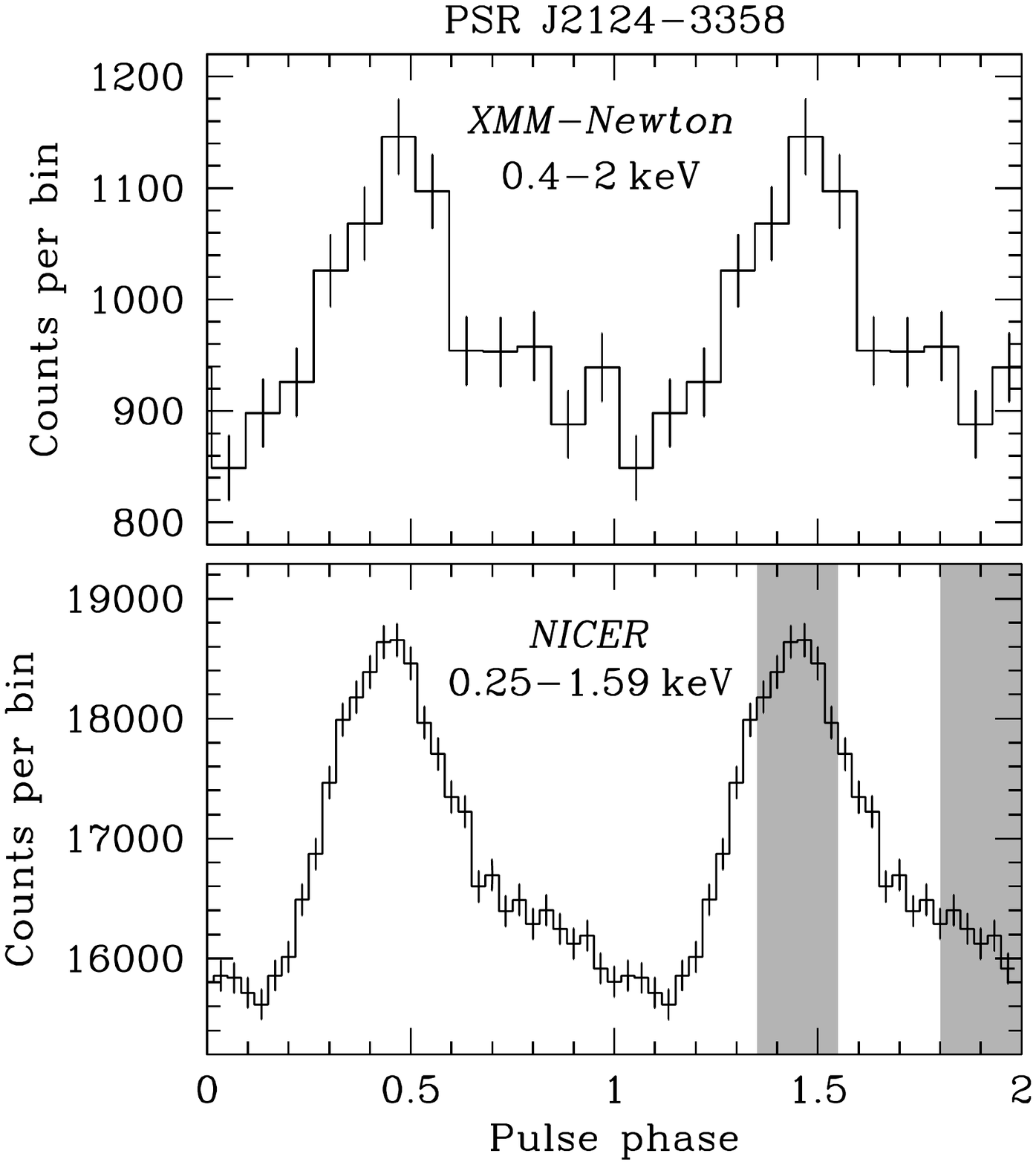}
\caption{Folded profiles of PSR~J0437--4715 (top left), PSR~J0030+0451 (top right), PSR~J1231$-$1411 (bottom left), and PSR~J2124$-$3358 (bottom right). In all instances, phase zero is determined by the radio ephemeris used for event folding. The upper panel for each MSP shows the previous best X-ray profile obtained with \textit{XMM-Newton} EPIC-pn,  with the exception of PSR~J1231$-$1411 for which no prior profile exists.  The grey bands mark the phase intervals used for the phase-selected spectroscopy described in Section~\ref{sec:spectroscopy}. Two rotational cycles are shown for clarity.}
\label{fig:msp_profiles}
\end{figure}

\begin{figure}
\centering
\includegraphics[clip,angle=0,width=0.47\textwidth]{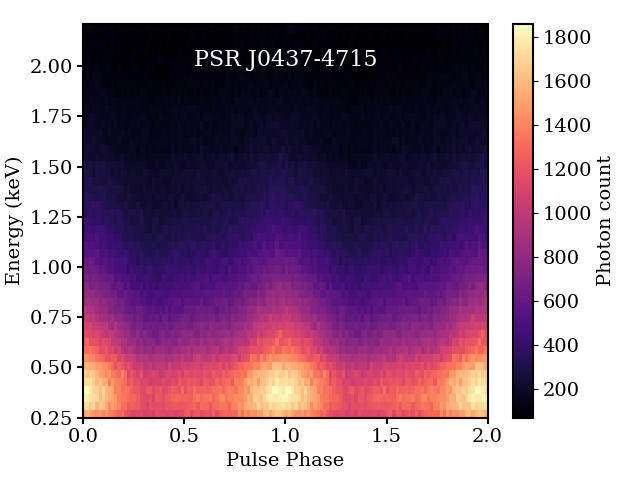}~~~
\includegraphics[clip,angle=0,width=0.47\textwidth]{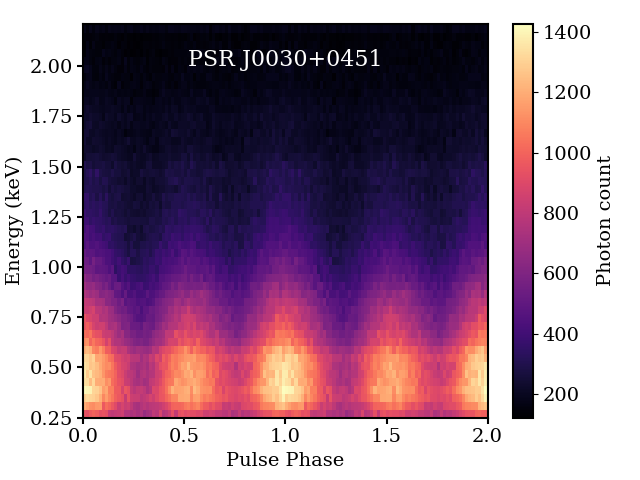}
\includegraphics[clip,angle=0,width=0.47\textwidth]{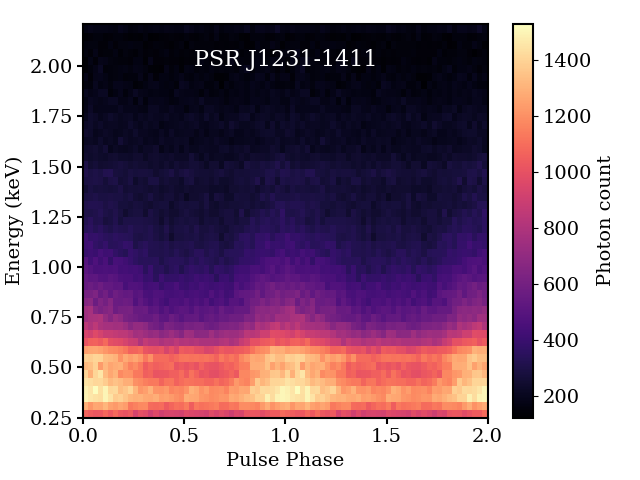}~~~
\includegraphics[clip,angle=0,width=0.47\textwidth]{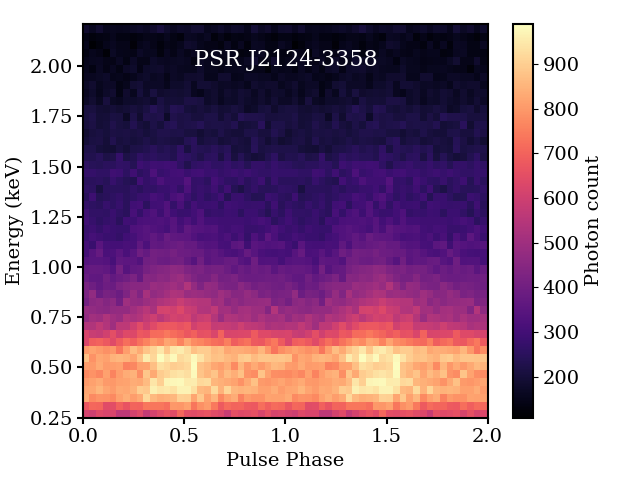}
\caption{Two-dimensional histograms of \textit{NICER} XTI counts versus pulse phase and photon energy for PSRs~J0437$-$4715, J0030$+$0451, J1231$-$1411, and  J2124$-$3358. The color bar shows the number of counts in each pixel. Two pulse phase cycles are shown for clarity. }
\label{fig:2d_profs}
\end{figure}

\subsection{PSR~J0437$-$4715}

PSR~J0437$-$4715 was discovered by \citet{johnston93} in the Parkes southern radio pulsar survey. At a distance of $156.79\pm0.25$ pc \citep{reardon16}, it is the nearest known MSP. It has properties typical of the Galactic population of MSPs, with a spin period $P=5.76$ ms and intrinsic spindown rate (after kinematic corrections) of $\dot{P}\equiv {\rm d}P/{\rm d}t = 1.37\times10^{-20}$ s s$^{-1}$, implying a surface dipole magnetic field strength $B\approx3\times10^8$ G, a characteristic age $\tau\approx4.9$\,Gyr, and spin-down luminosity $\dot{E}\approx3.8\times10^{33}$\,erg\,s$^{-1}$. The pulsar is bound to a $M=0.2$ M$_{\odot}$ helium-core white dwarf companion in a 5.74 day circular orbit \citep{bailyn93}.

PSR~J0437$-$4715 is the first radio MSP to be detected as a pulsed X-ray source with \textit{ROSAT} PSPC \citep{becker93}. Later, \textit{XMM-Newton} observations in timing mode were used to place constraints on the NS radius with pulse profile modeling (see \citealt{bogdanov13}).  Spectrally, the ultraviolet (UV) to hard X-ray emission ($\sim$ 0.01--20~keV) can be modeled with three thermal components and a non-thermal component \citep{durant12,bogdanov13,guillot16}.  The coldest thermal component describes the emission from the entire surface of the NS (excluding the hot spots), with a blackbody temperature $kT_{\rm BB}\sim 30$ eV. However, the size of the emission area is poorly constrained due to the limited sensitivity of X-ray instruments below $\approx$0.3~keV, and due to the limited coverage in the extreme UV regime where the Rayleigh-Jeans tail of this thermal component extends \citep{durant12}.  The two hotter thermal components are interpreted as originating from the hot spots, and are best modeled with NS atmosphere spectral components \citep{bogdanov13}. Finally, the non-thermal component is modeled with a simple power-law, with a best-fit photon index $\Gamma=1.50\pm0.25$, best constrained by \textit{NuSTAR} observations in the hard X-ray band \citep{guillot16}.  Timing analysis of these observations also revealed (3.7$\sigma$ detection) pulsations at the NS spin period in the 2--20~keV band \citep{guillot16}. Above 6~keV, where the power-law component dominates over the thermal emission by more than 2 orders of magnitude, the detection significance of these pulsations drops to $2.4\sigma$.  

PSR~J0437$-$4715 is located within 4\farcm18 of a bright Seyfert II active galactic nucleus (AGN), RX~J0437.4$-$4711 \citep{halpern96}.
There are also 11 other sources within 6\arcmin\ of the pulsar, identified through archival imaging observations with \textit{XMM-Newton}, and catalogued in the 3XMM-DR8 Catalog \citep{2016A&A...590A...1R}.  To minimize contamination due to the AGN and the other sources, we developed an optimization technique for the \nicer\ pointing. This method, described in Appendix~\ref{app:optimize}, finds the optimal pointing that maximizes the S/N from the pulsar by minimizing the total flux from nearby sources within the \nicer\ FOV.  For PSR~J0437$-$4715, the optimal pointing position is 1\farcm5 to the south-west of the pulsar, where the pulsar S/N is 16\% larger than for an on-source pointing. This is because, at the optimal pointing, the total contamination from other sources in the FOV amounts to 0.11\,s$^{-1}$, while it would be 0.82\,s$^{-1}$ (dominated by the AGN) if the pulsar were placed at the center of the FOV.

PSR~J0437$-$4715 has been observed regularly with \textit{NICER} since the mission's commissioning phase, with exposures starting on 2017 July 6 (ObsID 0060010101). Here, we present data obtained through 2019 March 12 (ObsID 2060010405). The folded pulse profile from \textit{NICER} based on 951 ks of clean exposure is shown in the upper left panel of Figure~\ref{fig:msp_profiles}. The strongest pulsed signal is found in the 0.25--1.85~keV range (at a 196.6$\sigma$ single trial significance). The asymmetric pulse profile is now seen with greater clarity compared to previous observations, especially the ``hump'' around phases 0.5--0.7, which in \citet{bogdanov13} is interpreted by invoking a second hot spot that is significantly displaced from the antipodal position relative to the primary spot. In the soft band (0.25--2~keV), where most of the source emission is found, the pulsations do not display any obvious changes as a function of energy (see upper left panel of Figure~\ref{fig:2d_profs}).

\subsection{PSR~J0030+0451} 

This solitary MSP was discovered at radio frequencies in the Arecibo drift scan survey \citep{lommen00} and is one of the nearest known MSPs ($D=325\pm9$ pc; \citealt{2018ApJS..235...37A}). Its spin period $P=4.87$ ms and intrinsic spindown rate $\dot{P}= 1.02\times10^{-20}$ s s$^{-1}$ imply a surface dipole magnetic field strength $B\approx2.7\times10^8$ G, a characteristic age $\tau\approx7.8$ Gyr, and a spin-down luminosity $\dot{E}\approx3\times10^{33}$\,erg\,s$^{-1}$. It was first detected in X-rays with \textit{ROSAT} \citep{2000ApJ...545.1015B}. Follow up observations with \textit{XMM-Newton} \citep{2002nsps.conf...64B,bogdanov09} showed that its emission spectrum in the 0.1--10~keV energy range is remarkably similar to that of PSR~J0437$-$4715, being well described by a predominantly thermal two-temperature model plus a faint hard tail evident above $\sim$3~keV. The pulsed emission in the 0.3--2~keV band is characterized by two broad pulses with pulsed fraction $\sim$60--70\%, consistent with a thermal origin of the X-rays, but only if the emission is significantly beamed such as may arise due to an atmosphere. 

The environment around PSR~J0030$+$0451 has many X-ray background sources. However, unlike the case of PSR~J0437$-$4715, these sources do not strongly contaminate the source counts in the \textit{NICER} observations. Our optimization method found that the pointing maximizing the S/N from the pulsar is 0\farcm25 in the north-east direction. However, the gain in S/N is $\sim0.1\%$, and this small offset pointing can be safely neglected for PSR~J0030$+$0451. Thus, for all observations of PSR~J0030+0451, \textit{NICER} was pointed at the pulsar position (see Appendix~\ref{app:optimize}).

The observations used for the parameter estimation analyses described in \cite{miller19} and \cite{riley19} were acquired over the period between 2017 July 24 (ObsID 1060020101) and 2018 December 9 (ObsID 1060020412). The \textit{NICER} pulse profile based on the resulting 1.936\,Ms of exposure in the 0.25--1.45~keV range (which yields the highest pulsed signal detection significance of 172.8$\sigma$) is shown in the upper right panels of Figure~\ref{fig:msp_profiles}. The high quality data reveal that the double peaked pulse profile retains its smoothness, as expected from surface thermal radiation from a NS, and confirm the significant difference in the amplitude of the two pulses and the depths of the two minima. In addition, as seen in Figure~\ref{fig:2d_profs}, the pulsed emission below $\sim2$~keV (where the source dominates above the background) remains unchanged in shape and phase alignment at all energies.

\subsection{PSR~J1231$-$1411}

This $P=3.68$ ms pulsar was discovered in a radio pulsar search campaign of unassociated \textit{Fermi} LAT sources with the Green Bank Telescope \citep{2011ApJ...727L..16R}. PSR~J1231$-$1411 is in a 1.86 day binary with a cool white dwarf companion. The pulsar dispersion measure implies a distance of 420 pc \citep{YMW16}. \textit{XMM-Newton} observations of this system have revealed a predominantly thermal spectrum \citep{2011ApJ...727L..16R}, reminiscent of PSRs~J0437$-$4715 and J0030+0451. With a 0.2--12~keV flux of $1.9\times10^{-13}$ erg cm$^{-2}$ s$^{-1}$, it is the third brightest thermally-emitting MSP and thus a well-suited target for \textit{NICER}. Prior to \textit{NICER} there were no X-ray observations of this MSP with sufficiently high time resolution to enable the detection of its X-ray pulsations. Timing and phase-averaged spectroscopic analyses were carried out by \cite{RayJ1231}, based on a subset of the \textit{NICER} data presented here. 

The environment around PSR~J1231$-$1411 has many X-ray background sources, but these sources are sufficiently faint to not contribute the majority of the expected counts during a \textit{NICER} observation. As for other pulsars, we use the 3XMM-DR8 Catalog \citep{2016A&A...590A...1R} to characterize these nearby sources, infer their expected \nicer{} count rates, and determine the optimal position to minimize their contribution to the background. For PSR~J1231$-$1411, we adopted a strategy to point at the pulsar position, since the gain in S/N would be just $\sim0.02\%$ for an optimal offset pointing of 0\farcm27 (see Appendix~\ref{app:optimize}).

The data presented here are based on observations with \textit{NICER} from 2017 June 26 to 2019 June 30, and include 440\,ks of additional clean exposure compared to \citet{RayJ1231}, for a total of 1.36\,Ms. The PSR~J1231$-$1411 \textit{NICER} profile in the 0.27--1.47~keV range (where the pulsed X-rays are detected at a maximum significance of 85$\sigma$) folded on the pulsar ephemeris from \cite{RayJ1231} is shown in the lower left panel of Figure~\ref{fig:msp_profiles}. The pulse morphology is distinct from those of the other MSPs considered here, in that it features a prominent broad main pulse and a much weaker (but statistically highly significant) secondary pulse. This is indicative of a significantly different hot spot configuration and/or viewing angle. An interesting feature of the main pulse is its slight asymmetry, with a trailing edge that is broader than the leading edge.  The pulsed emission from PSR~J1231$-$1411 is significantly softer (see Figure~\ref{fig:2d_profs} and Section~\ref{sec:spectroscopy}) compared to PSRs~J0437$-$4715 and J0030$+$0451, indicative of cooler polar caps.

\subsection{PSR~J2124$-$3358}

PSR~J2124$-$3358 is a nearby ($D=410\ud{90}{70}$ pc, \citealt{reardon16}), isolated MSP with a period $P=4.93$ ms \citep{bailes97}. It was first detected in X-rays by \textit{ROSAT} HRI \citep{becker99}. As the HRI provided no useful spectral information, only a total X-ray pulse profile was obtained, with pulsed fraction $\sim$33\%. PSR~J2124$-$3358 was observed with \textit{Chandra} ACIS-S for 30.2 ks and with the \textit{XMM-Newton} EPIC instrument for $\sim$70 ks \citep{zavlin06,hui06}. The spectrum of PSR~J2124$-$3358 is also adequately described by predominantly thermal emission with a 0.25--2 keV unabsorbed flux of $1.7\times10^{-13}$\,erg\,cm$^{-2}$\,s$^{-1}$.  
This MSP is also surrounded by diffuse X-ray emission due to a pulsar wind nebula or bow shock \citep{hui06,2017ApJ...851...61R}, contributing $\sim$4\% to the total emission (therefore adding to the background). 
In addition, a handful of nearby sources within $\sim 6\arcmin$ contribute to the observed \textit{NICER} count rate. The optimal offset pointing we chose is 1\arcmin\ to the south of the pulsar, resulting in a gain in S/N of 1.7\% compared to a pointing with the pulsar in the center of the FOV (see Appendix~\ref{app:optimize}).

The \textit{NICER} observations of PSR~J2124$-$3358 cover the period from 2017 June 26 
through 2019 June 30. 
The \textit{NICER} XTI pulse profile of PSR~J2124$-$3358 in the 0.25--1.59~keV band (where it is detected at a single trial significance of 39.4$\sigma$) is shown in the lower right panel of Figure~\ref{fig:msp_profiles}. The substantial improvement in photon statistics compared to the previous \textit{XMM-Newton} observation provide a much clearer sense of the pulse profile morphology. In particular, there is still no evidence for a distinct secondary pulse; instead, a trailing broad wing of the main pulse is now evident, resembling a mirrored version of the PSR~J0437$-$4715 profile. Although the signal-to-noise ratio of its pulse profile is lower compared to the other MSPs, PSR~J2124$-$3358 also does not show any clear profile evolution as a function of energy below $\sim$2~keV.

\begin{figure}
\centering
\includegraphics[clip,trim = 1cm 1cm 2cm 1cm, angle=0,width=0.45\textwidth]{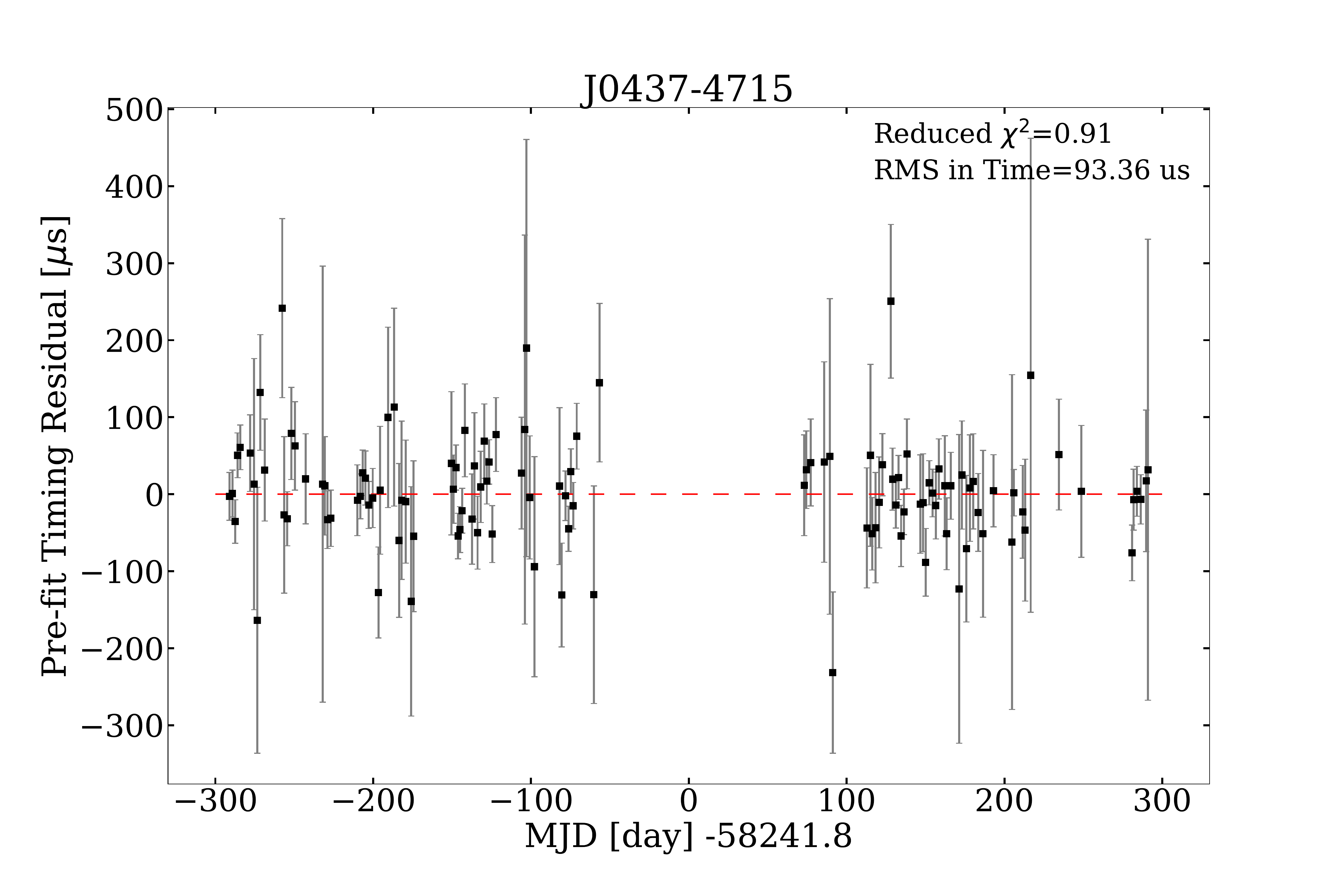}~~
\includegraphics[clip,trim = 1cm 1cm 2cm 1cm, angle=0,width=0.45\textwidth]{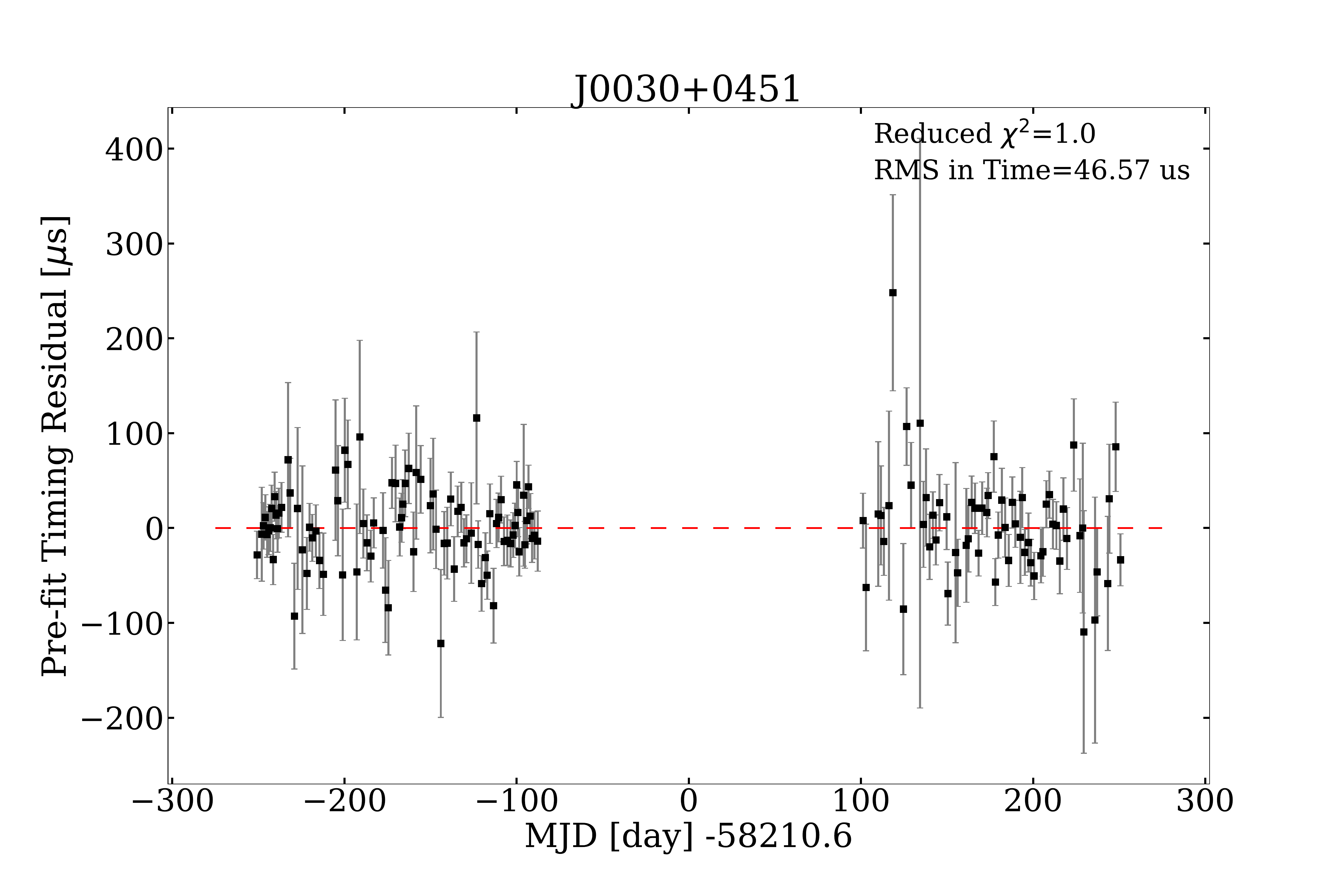}
\includegraphics[clip,trim = 1cm 1cm 2cm 1cm, angle=0,width=0.45\textwidth]{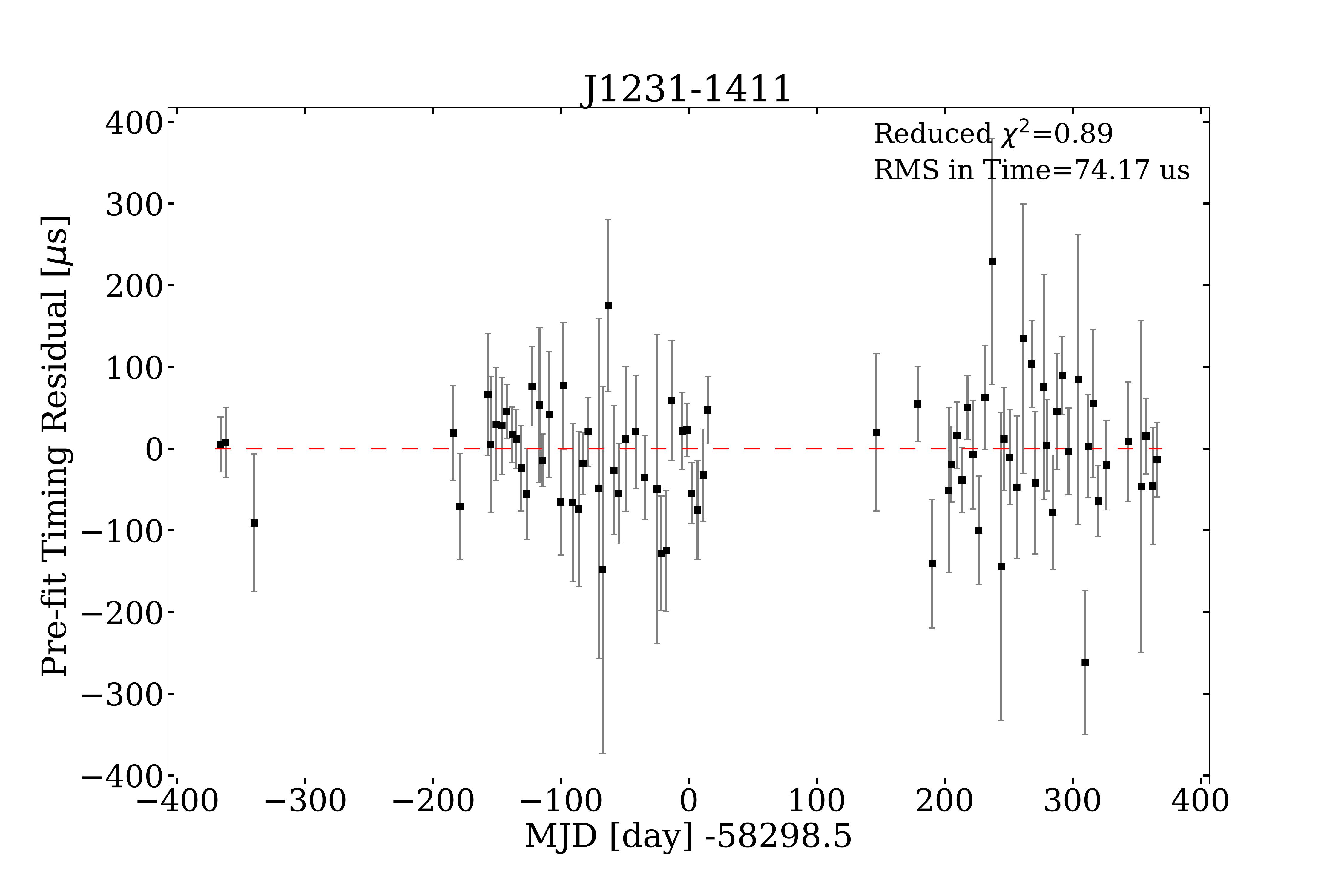}~~
\includegraphics[clip,trim = 1cm 1cm 2cm 1cm, angle=0,width=0.45\textwidth]{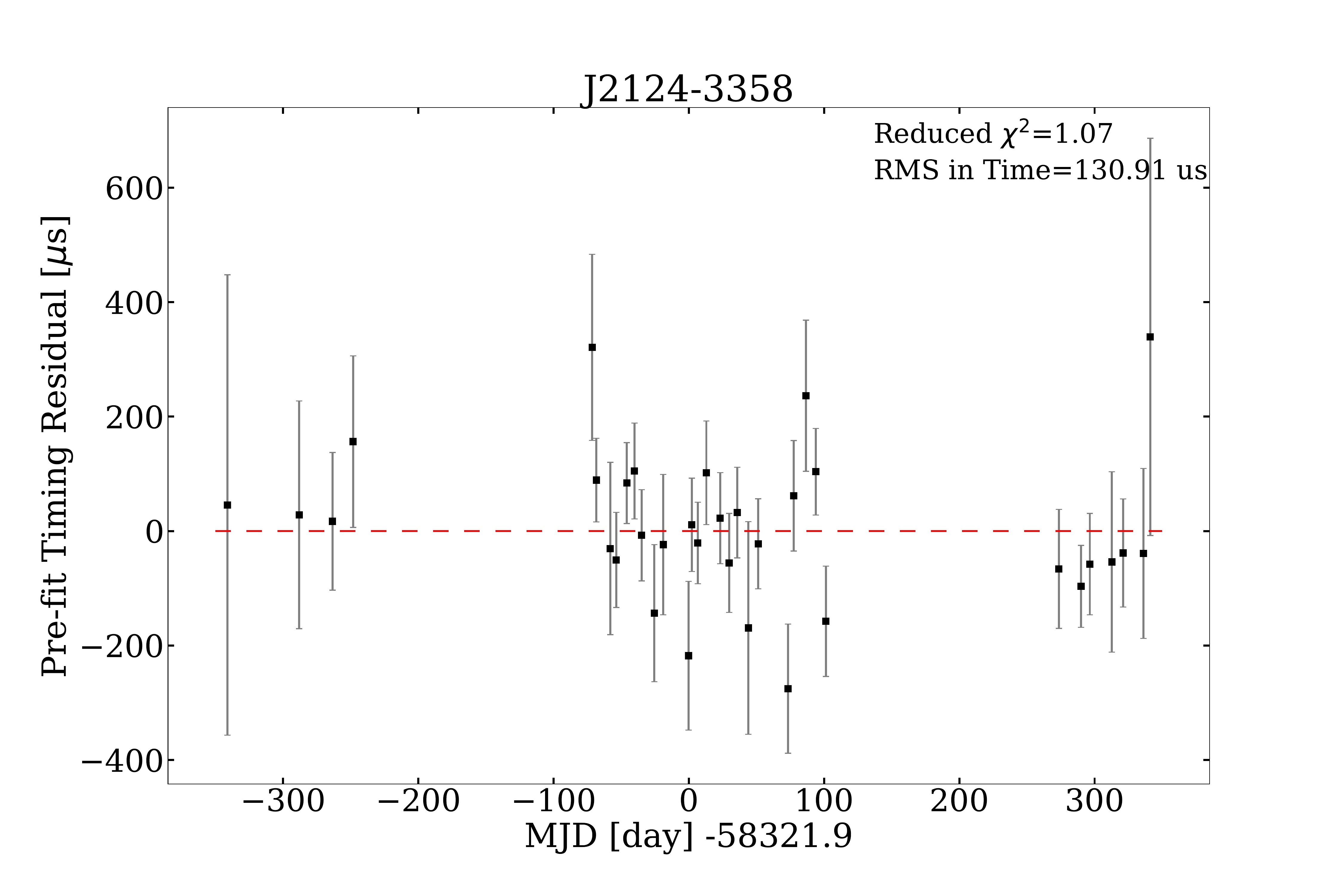}
\caption{Timing residuals for PSR~J0437$-$4715 (top left), PSR~J0030$+$0451 (top right), PSR~J1231$-$1411 (bottom left), and PSR~J2124$-$3358 (bottom right) for \textit{NICER} TOAs relative to the best available ephemeris for each pulsar.}
\label{fig:J0437_J0030_timing}
\end{figure}

\section{Event Folding, Long Term Timing and Flux Variability}
\label{sec:timing}

\subsection{Pulse Phase Assignment and Event Folding}

To obtain the folded \nicer\ pulse profiles using the entire span of available data, pulse phases were assigned to each event using two approaches:
\begin{itemize}

\item First applying the barycentric correction using the {\tt barycorr} tool in FTOOLS and assuming the DE421 JPL solar system ephemeris \citep{2009IPNPR.178C...1F}. Since the pulsars have measured proper motions,  for each observation, the position used for barycentering was computed based on the reference position and epoch and the measured proper motions in right ascension and declination. The resulting barycentered events were then folded with the pulsar ephemeris using the {\tt tempo2} ``photons'' plug-in. We note that this procedure is not strictly correct because it does not account for the pulsar parallax. Nevertheless, given the relatively small apparent motions of the MSPs over the \textit{NICER} data span, for the intended analysis this has a negligible effect ($\lesssim1$\,$\mu$s). 

\item Using the \texttt{photonphase} tool from the PINT pulsar timing package\footnote{\url{https://github.com/nanograv/pint}} and the \textit{NICER} orbit files (provided as one of the standard auxiliary products for each ObsID) to compute the transformation from the Terrestrial Time (TT) standard used for time tagging of \textit{NICER} events to Barycentric Dynamical Time (TDB) and to assign pulse phases based on an input pulsar ephemeris. In this approach, the proper motion and parallax given in the timing solution are explicitly taken into account. 

\end{itemize}
A comparison of the two methods shows differences at the level of $\lesssim 1$ $\mu$s in the form of a phase offset, which for the purposes of the analysis presented here is negligible. The results of the comparison of the two event folding approaches indicate that the procedures for assigning pulse phases to each event are reliable.

\subsection{Long Term Timing}
The energy-resolved pulse profiles used for the parameter estimation analyses aimed at constraining the neutron star $M$-$R$ relation and the dense matter EoS are based on \textit{NICER} observations carried out over time spans in the range 1.5--2 years. This does not present an issue due to the extraordinary rotational stability of MSPs and the availability of precise long-term timing solutions obtained from radio observations, which, when combined with the exquisite absolute timing capabilities of \textit{NICER}, permit the entire data set to be folded coherently at the pulsar period with negligible smearing of the pulse. 

To verify this assertion, we grouped the \textit{NICER} event data to produce time-of-arrival (TOA) measurements and compare them against the best available radio ephemerides.  In particular, for PSRs~J0437$-$4715 and J2124$-$3358, we use the Parkes Pulsar Timing Array timing solutions from \citet{reardon16}; for PSRs~J0030$+$0451, we use the ephemeris from the NANOGrav 11-year data set presented in \citet{2018ApJS..235...37A}, and for PSR~J1231$-$1411 we use the improved timing solution obtained by \citet{RayJ1231} using \textit{Fermi} LAT data. Each TOA was produced using 20 ks of effective observing time for PSRs~J0437$-$4715 and J0030+0451 using the same filtering procedure described previously, while requiring that each TOA span a time less than two days (172.8\,ks). An integration time of 50 ks per TOA was needed for  PSRs~J1231$-$1411 and J2124$-$3358 due to their dimmer nature and the maximum time span per TOA was relaxed to four days (345.6\,ks) and eight days (691.2\,ks), respectively. The TOAs were measured by fitting the data to a double Gaussian template, then following the maximum likelihood model described in \citet{RayJ1231}. Fitting was performed using the \texttt{photon\_toa} script from the NICERsoft package\footnote{Available from \url{https://github.com/paulray/NICERsoft}}. The residuals were produced using PINT, relative to the same radio timing solution used to create the profiles shown in Figures~\ref{fig:msp_profiles} and \ref{fig:2d_profs}. Figure~\ref{fig:J0437_J0030_timing} shows the \textit{NICER} timing residuals for PSRs~J0437$-$4715, J0030$+$0451, J1231$-$1411, and J2124$-$3358 when compared against their respective radio ephemerides. We note that no fitting of the timing parameters was involved in this comparison, i.e., we kept all pulsar parameters fixed and only fitted for a global spin phase offset. It is evident that, in general, the \textit{NICER} TOA measurements closely follow the radio timing solution as indicated by $\chi_{\nu}^2\approx 1$ in all cases. The resulting root-mean-square timing residuals are 44.01, 31.48, 53.07, and 100.13\,$\mu$s for PSRs~J0437$-$4715, J0030$+$0451, J1231$-$1411, and J2124$-$3358, respectively, which are at the expected level given the broad nature of the X-ray pulses. This provides assurance that folding the X-ray data over the entire observing span does not cause any smearing of the pulses that could negatively affect the desired $M$-$R$ measurement by distorting the intrinsic shape of the profiles.

\subsection{Long Term Flux Variability}
\label{sec:variability}

The rotation-powered MSPs considered here were chosen as targets for \textit{NICER} in part because they are not expected to exhibit any significant flux variability on timescales of years. Because the deep  \textit{NICER} data for each MSP span up to 2\,yr, we can examine the long-term behavior of the thermal X-ray flux in further detail. One complication is that the background emission local to \textit{NICER} (described in Section~\ref{sec:background}), including optical loading and energetic particles plus ambient (non-cosmic) radiation, exhibits long-term variability due to space weather, changing Sun angle, and the precession of the ISS orbit. Nevertheless, the pulsed component should be constant throughout, under the assumption that the polar cap radiation does not exhibit long-term temperature variations (caused, e.g., by variation in the return current).

To test for the presence of long-term variability, we divided each data set into a first half and a second half.  The dividing  point is determined by the halfway point in the counts, rather than the halfway point in time. For each half, we put the data into a form that has energy channels with 32 phase bins each.  We use energy channels 25 through 299 (0.25--2.99 keV) inclusive, i.e., 275 energy channels ($\Delta E=2.75$\,keV). We compare the halves of the data in the following way. We arbitrarily designated one half the ``data'' and the other half the ``model''. If there is no change in the underlying pulsed emission then we expect the  ``model'' to have the same fundamental folded profile as the "data", but the exposure time might not be the same and the contributions from other sources  (sky background, instrument noise, or space weather) could be different.  The zero of phase might also be different. Thus our ``model'' has one parameter per energy channel (a phase-independent background\footnote{The non-spot background (i.e. emission that does not originate from the surface hot-spots such as instrumental and sky backgrounds and non-thermal X-rays from the environment around the pulsar)  is treated in the same manner in the inference analyses for PSR~J0030+0451 presented in \cite{miller19} and \cite{riley19}.}), one parameter for the ratio of exposure  time between the ``model'' set and the ``data'' set, and one parameter for an  overall shift in phase. We optimized the match between the model and data using these  parameters. Using the optimized match, we then compute a $\chi^2$ between the ``model'' and the ``data'' for each phase-channel bin $\chi^2\,({\rm bin}) = ({\rm model}-{\rm data})^2/({\rm model}+{\rm data})$, which is effectively like assuming that the variance of the model in  each phase-channel bin is equal to the model in that bin, and similarly for the data, and that we can add the variances linearly to get the effective variance. Note that the sum of this pseudo-$\chi^2$ value is not expected to follow a true $\chi^2$ distribution exactly, because we are including the variance of the data from both halves.  Nonetheless, this offers a rough indicator of  whether the first and second halves are consistent with each other.

For PSRs~J0437$-$4715, J0030$+$0451, J1231$-$1411, and J2124$-$3358, we find $\chi^2$ values of  $8433.32$, $8841.09$, $8457.03$, and $8506.3$, respectively, for $8523$ degrees of freedom in all cases. The number of degrees of freedom is always $275\times 32-275-2$, i.e., the number  of phase-channel bins, minus the number of energy channels (because we  have a free background parameter per channel) minus an overall exposure time factor minus a phase shift.
For PSRs~J0437$-$4715, J1231$-$1411, and J2124$-$3358 the $\chi^2$ values are reassuringly small.  For PSR~J0030$+$0451, the formal probability of getting a $\chi^2$ that large or larger with that many degrees of freedom, if the model is correct, is 0.8\%.  However, given that we do not expect exactly a $\chi^2$ distribution, this is still consistent with no significant change between the first and second half for any of the four sources.
\begin{figure}
\centering
\includegraphics[width=0.44\textwidth]{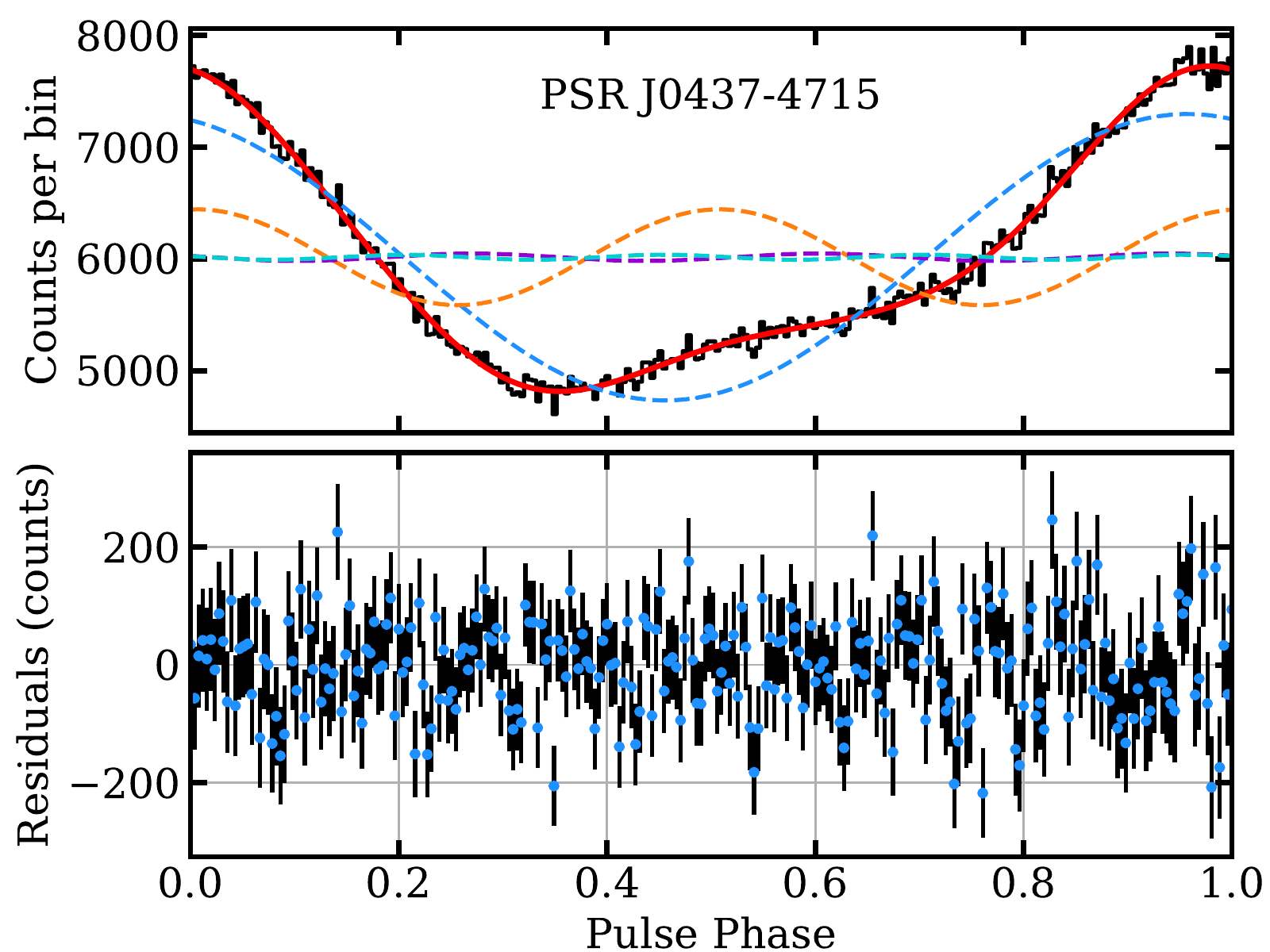}~~~~
\includegraphics[width=0.44\textwidth]{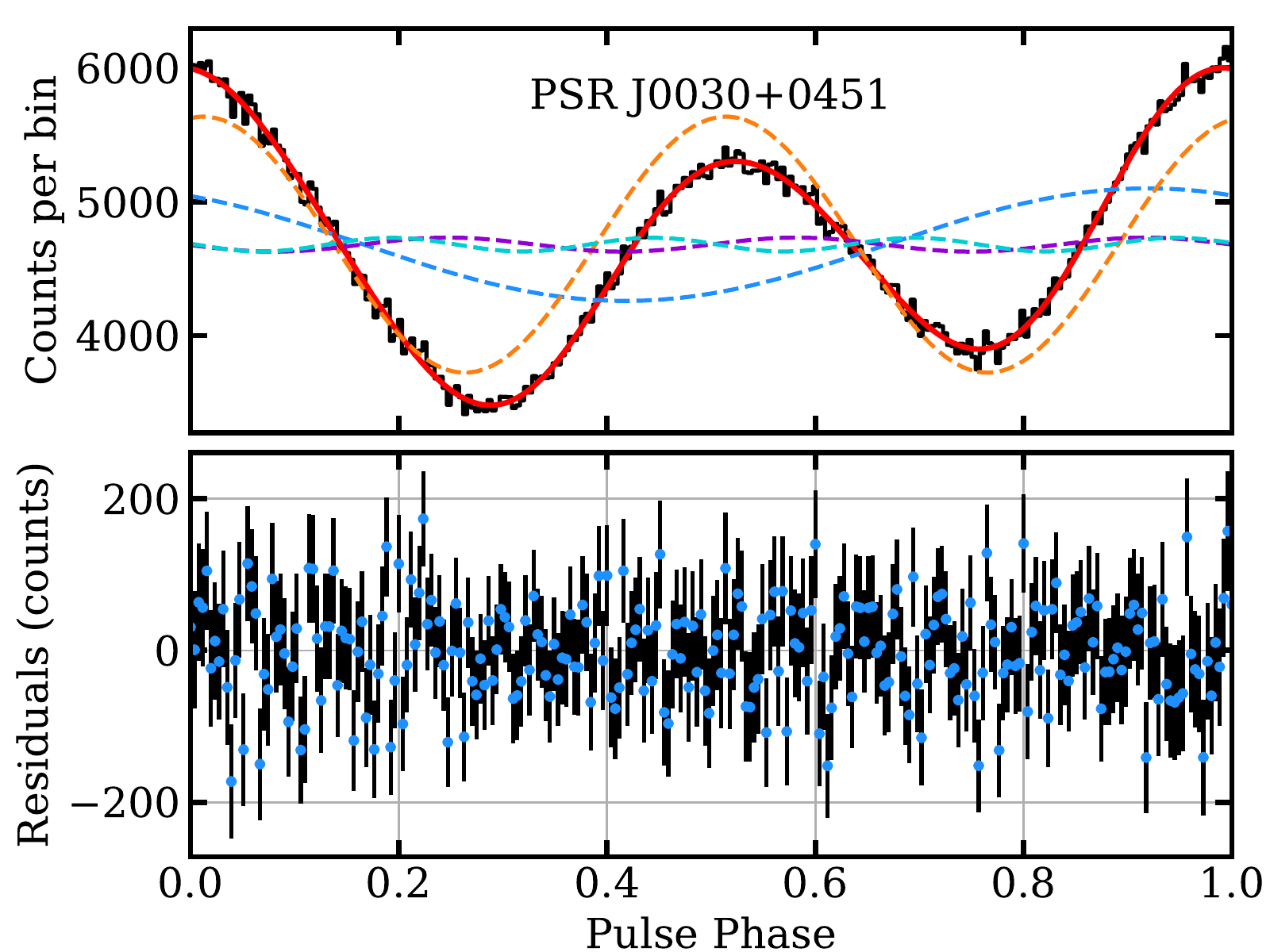}
\includegraphics[width=0.44\textwidth]{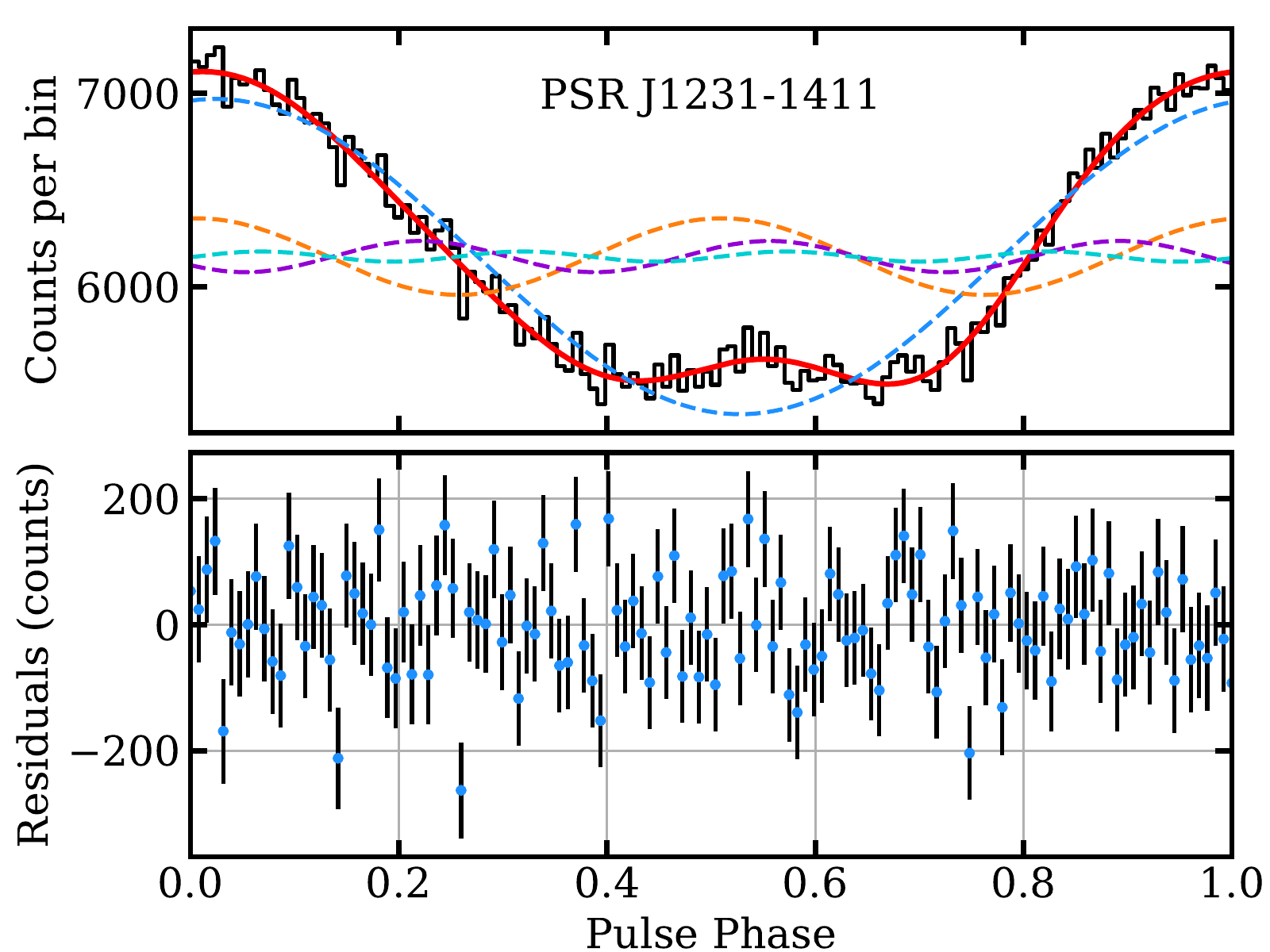}~~~~
\includegraphics[width=0.44\textwidth]{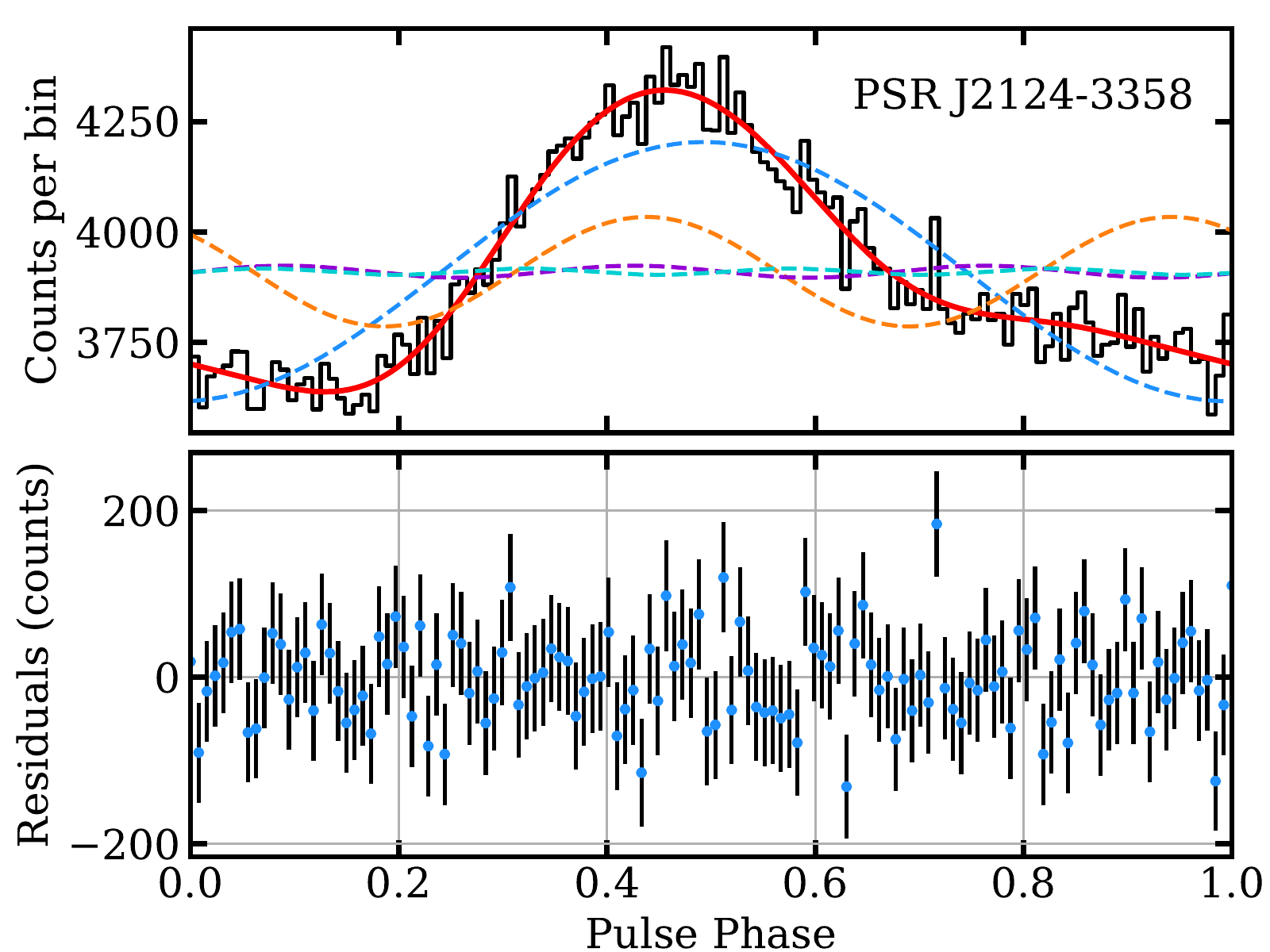}
\caption{The folded \textit{NICER} XTI profiles from Figure~\ref{fig:msp_profiles} but grouped in 256 (for PSRs~J0437$-$4715 and J0030$+$0451) or 128 phase bins (for PSRs~J1231$-$1411 and J2124$-$3358). The solid red lines show the best fit with a model of the profile constructed from the empirical Fourier coefficients, given by Equation~\ref{eq:fourier}, of the set of photon phases.   The dashed lines show the sinusoids corresponding to the first four harmonic components of the fit (blue, orange, purple, and cyan for the first, second, third, and fourth harmonic, respectively). The bottom panel for each pulsar shows the residuals from the fit.}
\label{fig:harmonics}
\end{figure}

\subsection{Fourier Decomposition of Pulse Profiles}
One property of the pulsed thermal X-ray emission that is important for NS $M$-$R$ constraints with the pulse profile modeling technique is the harmonic structure of the periodic signal.  The anisotropic beaming pattern of NS atmospheres \citep[e.g.,][]{zavlinetal96} causes the thermal pulsations to deviate from a sinusoidal shape, thus producing higher harmonics. Certain effects associated with the rapidly rotating NS such as occultation of the spot by the star, as well as Doppler boosting and aberration introduce extra harmonic content, thus providing useful information about  $M$ and $R$, which can be extracted through detailed modeling  \citep{miller98,weinberg01,2002ApJ...581..550M,poutanen06}.  As shown by \citet{miller15}, for stars with spin rates less than $\sim$300 Hz, the presence of the first (and higher) overtones of the spin frequency in the pulsed emission is due primarily to {\em i)} the non-isotropic beaming pattern of the radiation from the stellar surface and {\em ii)} the self-occultation of the hot spot(s) by the star. The harmonic content of the data also provides information regarding the optimal phase binning of the data, so that no useful information is lost by binning too coarsely.

With this in mind, we have examined the detection significance of the harmonics in the \textit{NICER} data as follows. Given a set of $N$ photons with computed phases $\varphi_i$ (in radians), we compute Fourier coefficients for harmonic $k$ as
\begin{equation}\label{eq:fourier}
\begin{split}
    c_k = \frac{2}{N}\sum_{i}\cos k \varphi_i \\
    s_k = \frac{2}{N}\sum_{i}\sin k \varphi_i.
\end{split}
\end{equation}
The Fourier coefficients define an analytic model for the pulse profile. We can compare it to the data by binning the event data and computing the model for each bin, then constructing $\chi^2$ as 
\begin{equation}
    \chi^2=\sum_{i}\frac{(b_i-m_i)^2}{b_i}
\end{equation}
where $b_i$ is the binned data and $m_i$ is the model prediction for that bin. If the fit is good, the residuals should be uncorrelated white noise (Poisson distributed, but effectively Gaussian-like for large numbers of counts per bin) and the reduced $\chi_{\nu}^2$ ($\chi^2$ per degree of freedom, d.o.f.) should be close to unity. Figure~\ref{fig:harmonics} shows the results of the fits of the model  profile constructed from the first four harmonically related sinusoids to the folded and binned profiles of PSRs~J0437$-$4715, J0030+0451, J2124$-$3358, and J1231$-$1411. For all MSPs, we find that four harmonics are sufficient to adequately describe the observed pulse profile, i.e., they yield $\chi_{\nu}^2\approx1.0$ and including higher harmonics does not significantly improve the fit. This is expected, since the thermal pulsations are relatively broad and smooth and the MSPs under consideration are not in a regime of NS spins where rapid rotation introduces strong higher harmonics. The residuals of the four-harmonic fits do not show any broad residuals,  statistically significant narrow features, or enhanced variance compared to what is expected.

\begin{figure}[t]
\begin{center}
\includegraphics[clip,angle=0,width=0.44\textwidth]{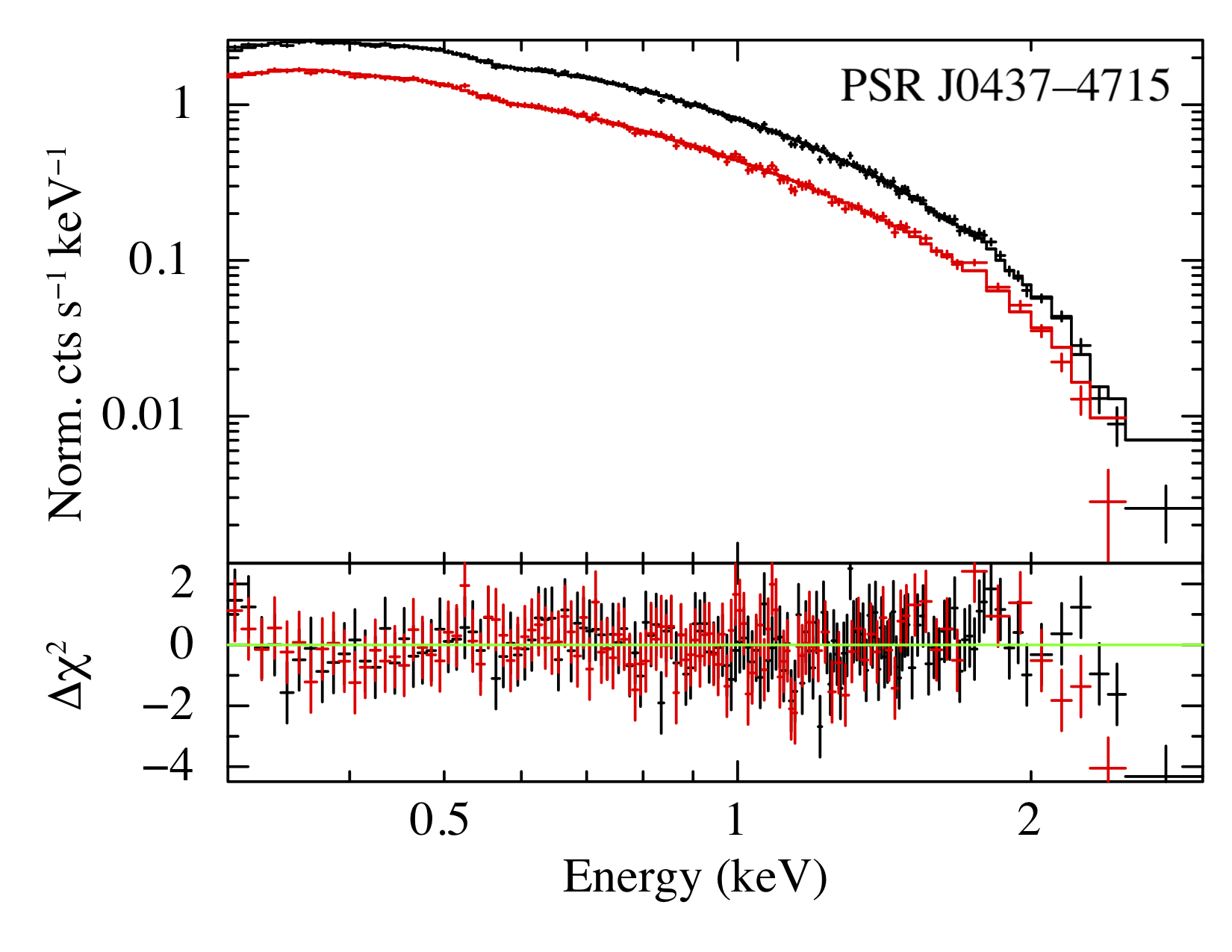}
\includegraphics[clip,angle=0,width=0.44\textwidth]{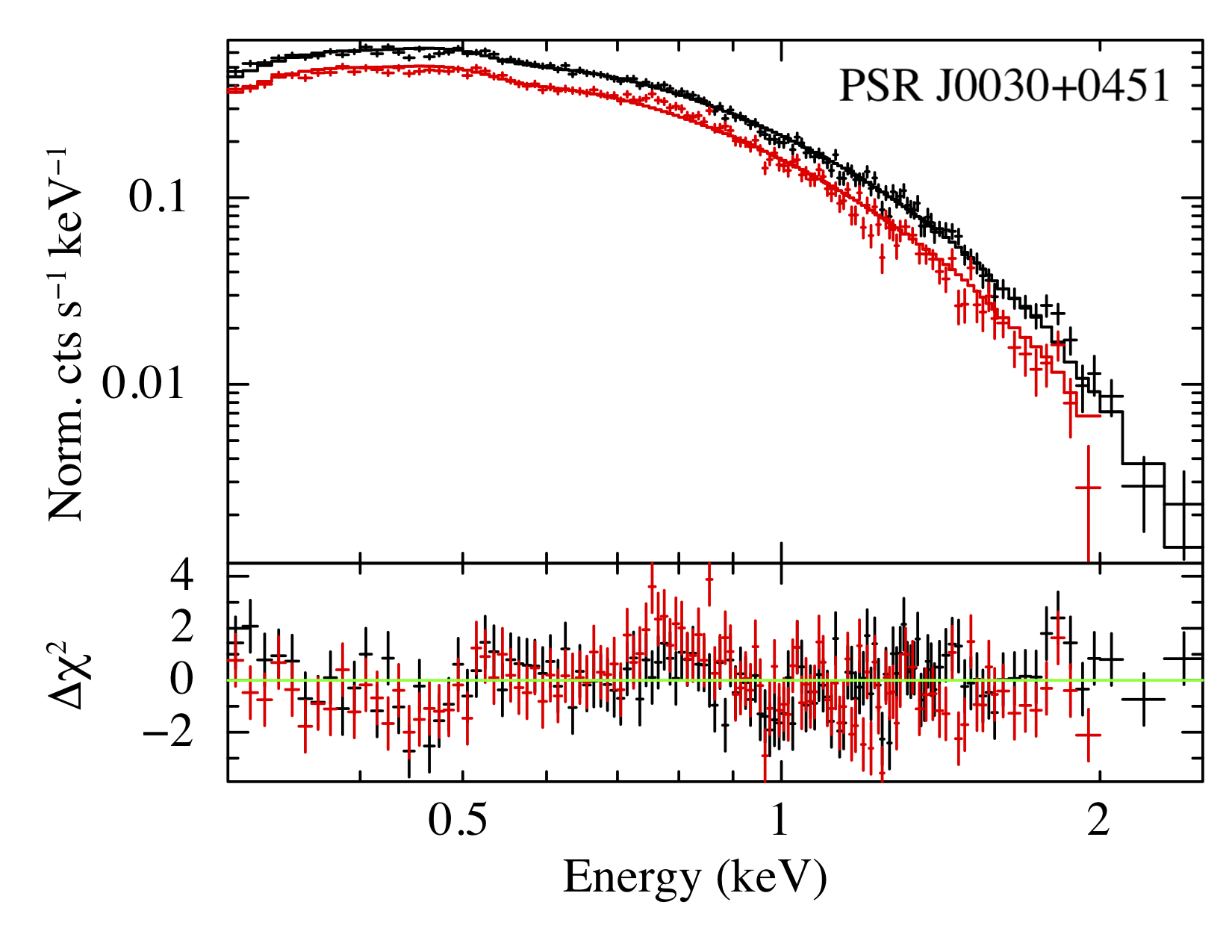}
\includegraphics[clip,angle=0,width=0.44\textwidth]{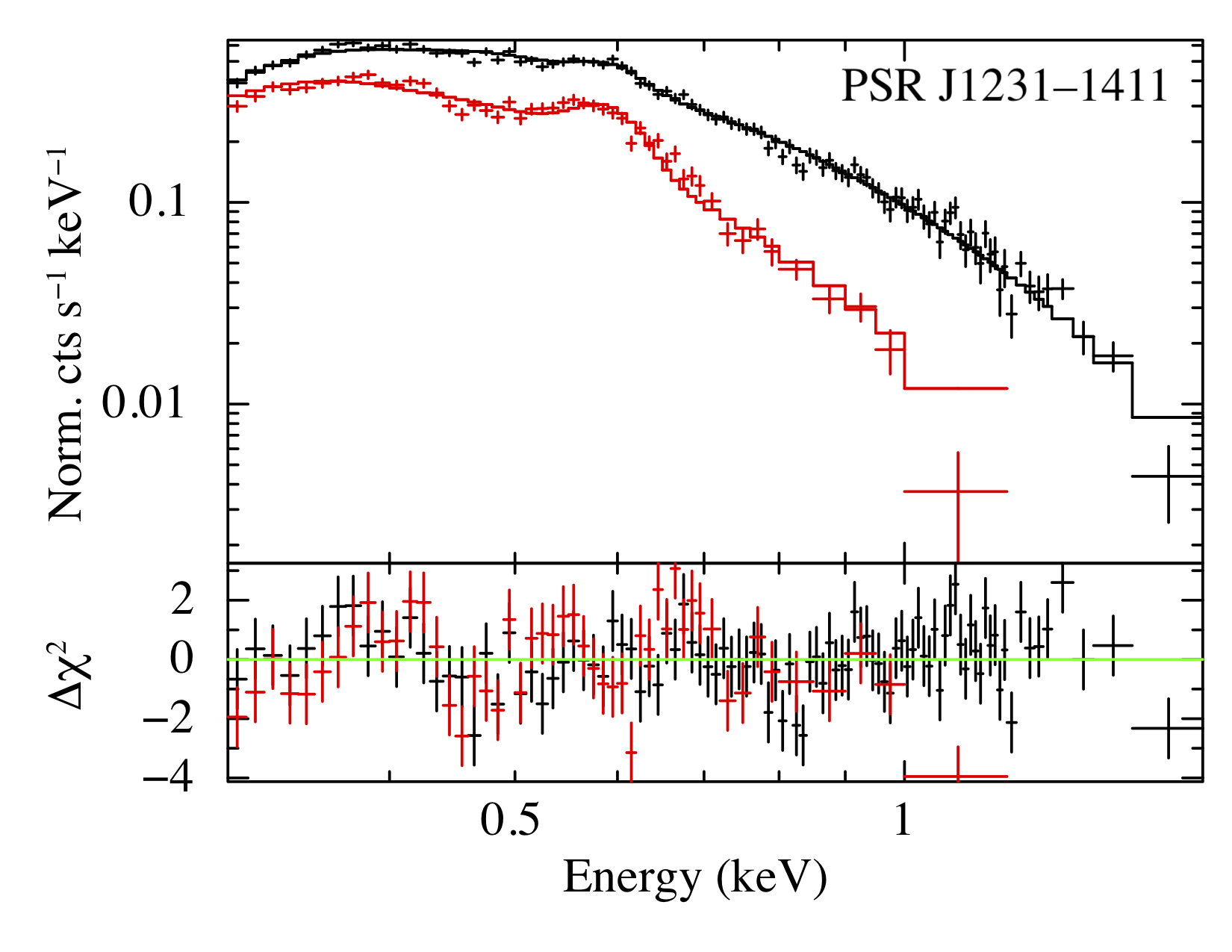}
\includegraphics[clip,angle=0,width=0.44\textwidth]{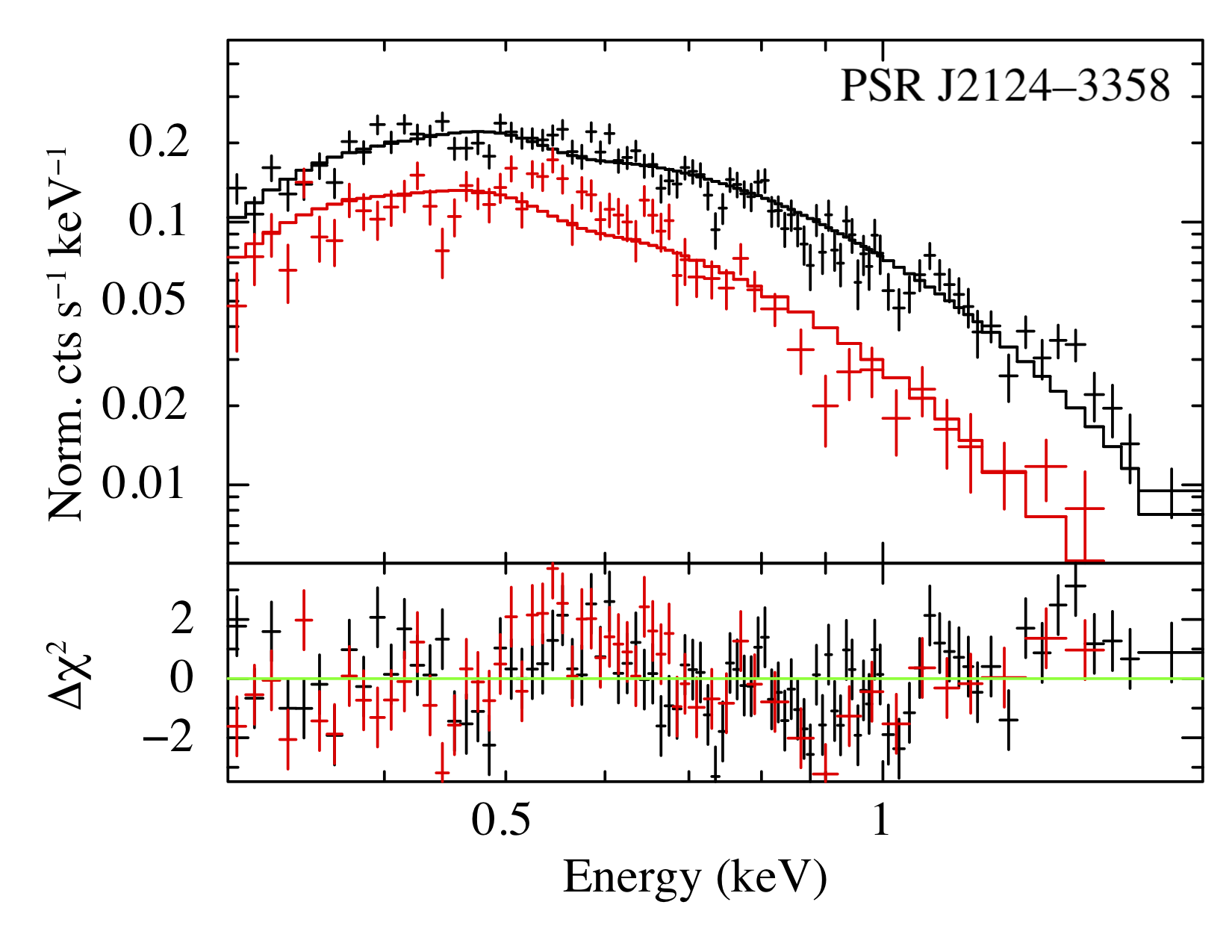}
\caption{Folded spectra of the two peaks of all four pulsars. In each panel, the dominant peak (peak 1) is represented in black and the secondary peak (peak 2) is in red. The bottom panel of each figure shows the residuals to the best-fit model in Tables~\ref{tab:spec0437}--\ref{tab:spec2124}.
}
\label{fig:spectra}
\end{center}
\end{figure}

\section{Phase-Selected Spectroscopic Analysis}
\label{sec:spectroscopy}

Previous analyses of the four MSPs have shown that their X-ray continua can be well described by a predominantly thermal spectrum (blackbody or neutron star atmosphere), with a requirement for more than one temperature for PSRs~J0437$-$4715 \citep{zavlin06,bogdanov13} and J0030$+$0451 \citep{bogdanov09}. For these two MSPs, above $\sim$3~keV, a power-law tail is seen in the spectrum.  It is important to note that there are limitations to standard phase-resolved spectroscopy, compared to the full phase-channel inference analyses used in \cite{miller19} and \cite{riley19}. For instance, the rotation of the NS and associated spin-phase flux averaging, as well as the relativistic effects and detailed geometry of the system, such as the location of the hot regions on the surface and the observer viewing angle are not taken into consideration. In addition, the NS atmosphere models available in XSPEC \citep{arnaud96} are constructed based on the assumption of uniform radiation from the entire surface of a NS but the emergent intensity of a NS atmosphere has a strong dependence on the emission angle \citep[see, e.g.,][]{zavlinetal96}. 
Moreover, even at pulse maximum the emitting region may not be viewed face-on.  As a consequence, the inferred temperatures and radii of the hot spots may significantly deviate from the true values.  Nevertheless, phase-resolved spectroscopy can still provide useful insight regarding the general properties of the surface radiation.

With these caveats in mind, we conducted phase-resolved spectroscopic analyses of these pulsars by selecting relatively narrow phase intervals around the peaks in the pulse profiles. Such phase selections (see Figure~\ref{fig:msp_profiles}) permit focusing on a single hot spot at a time, while also minimizing phase-averaging.  The spectra are therefore extracted within $\Delta\phi=0.2$ phase intervals around the peaks. For each pulsar, a model of the background is generated from the COR\_SAX and $K_{p}$ parameters of the GTIs used, with the space-weather based technique described in Section~\ref{sec:background}.  The event file filtering differs from that used for the full phase-channel analyses, and was optimized to produce the best match to the background model. For example, the minimum COR\_SAX value was set to 1.0 for PSR\,J2124$-$3358, PSR\,J1231$-$1411, and PSR\,J0437$-$4715, while it was maintained at 1.5 for PSR\,J0030$+$0451. These empirically produced background models provided a better match to the observed spectra above 3~keV, where the non-source emission dominates above the pulsar flux.

The spectrum used to model thermal emission is that of the NS hydrogen atmosphere model \texttt{nsatmos} \citep{heinkeetal06} in XSPEC, where 1, 2 or 3 such components were added when needed.  For each \texttt{nsatmos} component, we fit for the temperatures and normalizations (equivalent to the fraction of the total NS surface) and we fix the distances, masses, and radii, either to the known values when available or to canonical values.  Absorption due to the interstellar Galactic medium is modeled with \texttt{tbabs}, which employs the VERN cross-sections \citep{verner96} and WILM abundances \citep{2000ApJ...542..914W}. For each pulsar, the two peaks are fit simultaneously, keeping only the absorption parameter $N_{\rm H}$ tied between the two spectra. For PSR\,J1231$-$1411, a Gaussian component was added to account for excess of counts near the 0.57~keV \ion{O}{7} emission line, as observed in the phase-averaged analysis presented in \cite{RayJ1231}, and probably caused by Solar wind charge exchange or originating in the local hot bubble \citep{Kuntz2019}.  The other three pulsars do not exhibit such spectral features, likely due to weaker contamination from charge exchange along those lines of sight. Finally, we added a 3\% systematic to account for uncertainties in the background modeling.  We note that in all cases but PSR~J0437$-$4715, the backgrounds represent $\gtrsim 50\%$ of the total extracted counts in the bands used, and as much as $\sim80\%$ for PSR~J2124$-$3358. Uncertainties in the modelling of the background spectra can have important effects on the pulsar spectral analyses (see Section~\ref{sec:background} for details).  The results for each target are presented in the following subsections and the best-fit models to the spectral data are displayed in Figure~\ref{fig:spectra}.

\subsection{PSR~J0437$-$4715}
In the analysis of PSR~J0437$-$4715, the following parameters of each \texttt{nsatmos} spectral component are kept fixed: $M=1.44$~M$_{\odot}$ and $D=156.79$~pc, since they are precisely and independently measured \citep{reardon16}. Neither single-\texttt{nsatmos} nor double-\texttt{nsatmos} spectral models describe the data well (\Chisq= 6806.81 for 258 d.o.f., and \Chisq = 394.17 for 254 d.o.f., respectively).  As was observed in previous phase-averaged analyses of this pulsar, three thermal components are necessary to model the observed emission. The coldest \texttt{nsatmos} component is expected to emerge from the entire surface of the NS \citep{durant12,bogdanov13,guillot16,gonzalez19}, and is therefore expected to be visible at all phases.  Therefore, each spectrum studied here (one for each peak) likely displays the emission from a two-temperature polar cap in addition to the emission from the entire surface, the latter assumed to be the same for each peak (although with different effective emission areas).

For this reason, we choose to keep the parameters of the cold \texttt{nsatmos} surface component tied between the spectra of the two peaks, except for their normalizations, which are constrained to be equal to $1$ minus the normalizations of the other two \texttt{nsatmos} components for the same peak.  As a result, the cold \texttt{nsatmos} normalizations are not degenerate with the NS radius, which can then be a free parameter.  We can therefore fit the NS radius at the same time as the parameters from the hot polar caps emission. Although this does not have the robustness of a full phase-energy resolved analysis, the results, presented in Table~\ref{tab:spec0437}, can be informative.

We find that the \texttt{nsatmos2} (mid-temperature) component has comparable temperatures in both peaks. However, the \texttt{nsatmos3} (high-temperature) component appears hotter in the second (less-prominent) peak.  As expected from their contributions to the total pulse profile, the second peak has smaller \texttt{nsatmos2} and \texttt{nsatmos3} normalizations, indicating a polar cap that is either smaller than that of peak 1, or is viewed at a larger angle by the observer.  This ought to be clarified from detailed modeling of the phase-energy resolved data (as described in \citealt{miller19} and \citealt{riley19}), which will be presented in subsequent publications).  The best-fit temperature and radius of the coldest \texttt{nsatmos} component (\texttt{nsatmos1}) have to be interpreted with care. Indeed, its contribution to the total spectrum is minimal (as it peaks outside the \nicer{} band) and it is only constrained at the lowest energies where contamination from optical loading may still be present, even when considering Sun angles $>80^{\circ}$ (see Section~\ref{sec:background}). We note that in this spectroscopic analysis the NS radius is not well constrained primarily because it does not take into account the full three-dimensional geometry and rotation of the system. Therefore, it is not nearly as sensitive to the NS radius as a full inference analysis that considers the energy-dependent beaming pattern of the NS atmosphere, gravitational bending of light, viewing geometry, and stellar rotation. In addition, due to the way the \texttt{nsatmos} model is parameterized, the NS radius and the flux normalization are covariant, which increases the uncertainty in $R$.


\begin{deluxetable}{llcc}
\tablecaption{Results of the \nicer{} spectral analysis for PSR~J0437$-$4715.  \label{tab:spec0437}}
\tablehead{
\colhead{Component} & \colhead{Parameter} & \colhead{\tt Peak 1} & \colhead{\tt Peak 2}
}
\startdata
{\tt tbabs}    & $N_{\rm H}$ ($\ee{20}\percmsq$) & \multicolumn{2}{c}{0.08\ud{0.20}{0.08}} \\
\hline
{\tt nsatmos1}  & $\log{T_{\rm eff}}$ & \multicolumn{2}{c}{$5.26\ud{0.05}{0.18}$} \\
               & $M$ (M$_{\odot}$)    & \multicolumn{2}{c}{(1.44)} \\
               & $R$ (km)             & \multicolumn{2}{c}{15.3\ud{2.0}{1.6}} \\
               & $D$ (pc)             & \multicolumn{2}{c}{(156.79)} \\
               & Norm.                & $1.0-(N_2+N_3)$ & $1.0-(N_2+N_3)$ \\
\hline
{\tt nsatmos2}  & $\log{T_{\rm eff}}$ & $5.71\ud{0.05}{0.05}$ & $5.74\ud{0.05}{0.05}$\\
               & Norm. $N_2$          & $0.036\ud{0.018}{0.012}$ & $0.021\ud{0.012}{0.008}$\\
\hline
{\tt nsatmos3}  & $\log{T_{\rm eff}}$ & $6.23\ud{0.02}{0.01}$ & $6.29\ud{0.03}{0.02}$ \\
               & Norm. $N_3$          & $0.00035\ud{0.00011}{0.00008}$ & $0.00009\ud{0.00003}{0.00002}$\\
\hline
\multicolumn{2}{c}{Count rate (s$^{-1}$)} & $1.509$ & $0.887$ \\
\multicolumn{2}{c}{$F_{\rm 0.3-2.0\,keV}$ ($\ee{-13}\cgsflux$) } & $15.78\pm0.05$ & $9.28\pm0.05$  \\
\multicolumn{2}{c}{$L_{\rm 0.3-2.0\,keV}$ ($\ee{30}\cgslum$) } & $4.64\pm0.01$ & $2.73\pm0.01$  \\
\hline
\multicolumn{2}{c}{\Chisqred\ (d.o.f.)} & \multicolumn{2}{c}{0.84 (252)}
\enddata
\tablecomments{Values in parentheses are kept fixed. Due to the complex shape of the parameter space, errors were estimated from a Markov Chain Monte Carlo within XSPEC (500000 iterations, with 100 walkers). Reported values correspond to the 50\% quantiles, and the upper/lower uncertainties are the 5\% and 95\% quantiles so that they represent the 90\% credible intervals.  A multiplicative constant with fixed value of 0.95 was used to account for the 1.5\arcmin\ offset pointing (see Figure~\ref{fig:vignetting}).}
\end{deluxetable}

\subsection{PSR~J0030$+$0451}
In the spectral analysis of PSR~J0030$+$0451, we fix the NS radius and mass parameters to $R=12.7$\,km and $M=1.34$\,M$_{\odot}$ (from \citealt{riley19}), as well as the distance $d=325$~pc---but we note that the fit is not very sensitive to small variations of $M$ and $R$, which can be absorbed in adjustments of the other parameters. We obtain a marginally acceptable fit with $\Chisqred=1.35$ (for 250 d.o.f.).  The temperature and normalizations of the two peaks are reported in Table~\ref{tab:spec0030}. Using the nominal best fit values for $R=13.02$\,km and $M=1.44$\,M$_{\odot}$  from \citet{miller19} produces similar results.

Adding another \texttt{nsatmos} component (with smaller temperature) is not strictly required by the data\footnote{A simulation, using \texttt{simftest} in XSPEC finds a probability of 0.005 (not quite $3\sigma$) that the additional component is required.}. Therefore, we report here the fit with only one thermal component.  We find temperatures that are colder than those reported in \cite{miller19} and \cite{riley19}:  $\log{T_{\rm eff}}\sim 6.11$ for both polar caps, Peak 1 corresponding to the crescent/oval and Peak 2 to the circular spot in the favored models reported in those papers.  The difference between the temperatures of Table~\ref{tab:spec0030} and those in \cite{miller19} and \cite{riley19} can be attributed to a combination of the present analysis considering spectra averaged over 0.2 in pulse phase, not fully accounting for  the viewing geometry of the system and all relevant general and special relativistic effects, and the use of atmosphere model spectra integrated over all angles.


\begin{deluxetable}{llcc}
\tablecaption{Results of the \nicer{} spectral analysis for PSR~J0030$+$0451. \label{tab:spec0030}}
\tablehead{
\colhead{Component} & \colhead{Parameter} & \colhead{\tt Peak 1} & \colhead{\tt Peak 2}
}
\startdata
{\tt tbabs}    & $N_{\rm H}$ ($\ee{20}\percmsq$) & \multicolumn{2}{c}{0.03\ud{0.25}{0.03}} \\
{\tt nsatmos}  & $\log{T_{\rm eff}}$ & $6.036\ud{0.004}{0.007}$     & $6.014\ud{0.005}{0.007}$ \\
               & $M$ (M$_{\odot}$)            & \multicolumn{2}{c}{(1.34)} \\
               & $R$ (km)              & \multicolumn{2}{c}{(12.7)} \\
               & $D$ (pc)              & \multicolumn{2}{c}{(325)} \\
               & Norm.                & $0.0055\ud{0.0005}{0.0002}$ & $0.0052\ud{0.0005}{0.0003}$\\
\hline
\multicolumn{2}{c}{Count rate (s$^{-1}$)} & $0.405$ & $0.312$ \\
\multicolumn{2}{c}{$F_{\rm 0.3-2.0\,keV}$ ($\ee{-13}\cgsflux$) } & $3.78\pm0.02$ & $2.92\pm0.02$  \\
\multicolumn{2}{c}{$L_{\rm 0.3-2.0\,keV}$ ($\ee{30}\cgslum$) } & $4.78\pm0.02$ & $3.69\pm0.02$  \\
\hline
\multicolumn{2}{c}{\Chisqred\ (d.o.f.)} & \multicolumn{2}{c}{1.35 (250)}
\enddata
\tablecomments{Values in parentheses are kept fixed. All errors reported are at 90\% confidence.}
\end{deluxetable}

\subsection{PSR~J1231$-$1411}

A phase averaged spectral analysis of a 916~ks subset of the data studied in the present paper has been performed in the article summarizing the discovery of pulsations from this pulsar \citep{RayJ1231}. The authors concluded that a double-blackbody model (with temperatures 44 and 133~eV) or a single NS hydrogen atmosphere model with effective temperature 51~eV was required to fit the data.  Here we analyze the enlarged data set for PSR~J1231$-$1411, i.e., 1320~ks of good time after optimal filtering for the spectral analysis.  Since the two peaks are separated into two spectra, a single \texttt{nsatmos} component is fitted to each, assuming $M=1.4$\,M$_{\odot}$ and $R=11$\,km. The assumed value of $R$ is based on existing measurements from quiescent NS X-ray binaries \citep[see, e.g.,][and references therein]{2018MNRAS.476..421S}; choosing a different value does not change the main conclusions of the analysis. As with the other pulsars, only the absorption parameter $N_{\rm H}$ is tied between the two peaks. Furthermore, as was found in \cite{RayJ1231}, an \ion{O}{7} line feature is necessary to fit the thermal spectrum of this pulsar (the line parameters are also tied between the two spectra).  We find \texttt{nsatmos} effective temperatures corresponding to $61$~eV and $37$~eV for the two peaks, respectively.  Results are presented in Table~\ref{tab:spec1231}.


\begin{deluxetable}{llcc}
\tablecaption{Results of the \nicer{} spectral analysis for PSR~J1231$-$1411. \label{tab:spec1231}}
\tablehead{
\colhead{Component} & \colhead{Parameter} & \colhead{\tt Peak 1} & \colhead{\tt Peak 2}
}
\startdata
{\tt tbabs}    & $N_{\rm H}$ ($\ee{20}\percmsq$) & \multicolumn{2}{c}{$1.7\pm0.3$} \\
{\tt Gaussian} & $E_{\rm G}$ (keV)      & \multicolumn{2}{c}{$0.590\ud{0.005}{0.001}$}\\
               & $\sigma_{\rm G}$ (keV) & \multicolumn{2}{c}{$(1.5\pm0.2)\tee{-5}$}\\
               & Norm ($\ee{-5}$ ph $\percmsq\persec$) & \multicolumn{2}{c}{$(1.3\pm0.2)\tee{-5}$}\\
{\tt nsatmos}  & $\log{T_{\rm eff}}$ & $5.85\ud{0.01}{0.01}$     & $5.63\ud{0.02}{0.02}$ \\
               & $M$ (M$_{\odot}$)  & \multicolumn{2}{c}{(1.4)} \\
               & $R$ (km)              & \multicolumn{2}{c}{(11.0) }\\ 
               & $D$ (pc)              & \multicolumn{2}{c}{(420)} \\
               & Norm.                & $0.075\ud{0.014}{0.011}$ & $0.48\ud{0.13}{0.10}$\\
\hline
\multicolumn{2}{c}{Count rate (s$^{-1}$)} & $0.27$ & $0.13$ \\
\multicolumn{2}{c}{$F_{\rm 0.3-2.0\,keV}$ ($\ee{-13}\cgsflux$) } & $2.56\pm0.02$ & $1.28\ud{0.01}{0.03}$  \\
\multicolumn{2}{c}{$L_{\rm 0.3-2.0\,keV}$ ($\ee{30}\cgslum$) } & $5.40\pm0.05$ & $2.70\ud{0.02}{0.06}$  \\
\hline
\multicolumn{2}{c}{\Chisqred (d.o.f.)} & \multicolumn{2}{c}{1.49 (141)}
\enddata
\tablecomments{Values in parentheses are kept fixed. All errors reported are at 90\% confidence.}
\end{deluxetable}

\subsection{PSR~J2124$-$3358}
PSR~J2124$-$3358 is the faintest of the four pulsars presented here. The results of its spectral analysis are somewhat uncertain and dependent on the reliability of the background model. Indeed, the pulsar count rates represent only 24\% and 15\% of the total counts for the spectra of the dominant peak and secondary peak, respectively, in the 0.3--1.5~keV band. We again assume fixed values of $M=1.4$\,M$_{\odot}$ and $R=11$\,km.  The analysis, similar to the other pulsars described above, finds the dominant peak with a higher temperature but smaller size (although not significantly) than the secondary peak (Table~\ref{tab:spec2124}).  These findings are qualitatively consistent with those of \cite{bogdanov08}, if the two thermal components used in that work are assumed to originate from the two peaks in the pulse profile as studied here.

Finally, we note that the cold emission from the entire pulsar surface ($T_{\rm BB}=(0.5-2.1)\tee{5}$~K; see \citealt{Rangelov17}) measured in the far-UV with the \textit{Hubble Space Telescope} remains mostly outside the \nicer{} band, and therefore cannot be detected (unlike for PSR~J0437$-$4715).


\begin{deluxetable}{llcc}
\tablecaption{Results of the \nicer{} spectral analysis for PSR~J2124$-$3358. \label{tab:spec2124}}
\tablehead{
\colhead{Component} & \colhead{Parameter} & \colhead{\tt Peak 1} & \colhead{\tt Peak 2}
}
\startdata
{\tt tbabs}    & $N_{\rm H}$ ($\ee{20}\percmsq$) & \multicolumn{2}{c}{3.1\ud{0.8}{0.8}} \\
{\tt nsatmos}  & $\log{T_{\rm eff}}$ & $5.99\pm0.02$            & $5.88\pm0.03$ \\
               & $M$ (M$_{\odot}$)    & \multicolumn{2}{c}{(1.4)} \\
               & $R$ (km)              & \multicolumn{2}{c}{(11.0)} \\
               & $D$ (pc)              & \multicolumn{2}{c}{(410)} \\
               & Norm.                & $0.008\ud{0.003}{0.002}$ & $0.014\ud{0.006}{0.005}$\\
\hline
\multicolumn{2}{c}{Count rate (s$^{-1}$)} & $0.130$ & $0.063$ \\
\multicolumn{2}{c}{$F_{\rm 0.3-2.0\,keV}$ ($\ee{-13}\cgsflux$) } & $1.27\ud{0.01}{0.04}$ & $0.62\ud{0.01}{0.03}$  \\
\multicolumn{2}{c}{$L_{\rm 0.3-2.0\,keV}$ ($\ee{30}\cgslum$) }   & $2.54\ud{0.02}{0.06}$ & $1.26\ud{0.02}{0.06}$  \\
\hline
\multicolumn{2}{c}{\Chisqred\ (d.o.f.)} & \multicolumn{2}{c}{1.92 (142)}
\enddata
\tablecomments{Values in parentheses are kept fixed. All errors reported are at 90\% confidence. A multiplicative constant with fixed value of 0.98 was used to account for the 1\arcmin\ offset pointing (see Figure~\ref{fig:vignetting}).}
\end{deluxetable}

\section{Conclusions}\label{sec:conclusions}
In this paper, we presented the deep \textit{NICER} data sets of PSRs~J0437$-$4715, J0030$+$0451, J1231$-$1411, and J2124$-$3358,  the four rotation-powered MSPs we have selected for analysis aimed at constraining the NS $M-R$ relation and the dense matter EoS. The data were reduced using the best available tools and data cleaning techniques the \textit{NICER} team has developed. The filtered and phase folded event data set for each MSP is provided as supplemental material to this article.   The substantial increase in photon statistics of these data compared to previous observations of these targets enables better characterization of their pulsed emission.  We examined the timing behavior, long term variability, Fourier domain characteristics, and spectral properties of the event data. 

We confirm that the superb absolute timing capabilities of \textit{NICER} enable folding of the event data, spanning 1.5--2 yr, coherently with the radio ephemerides with virtually no pulse smearing. We find that the pulsed emission does not exhibit significant changes over the observing span, as expected from polar cap emission from rotation-powered MSPs. A Fourier decomposition of the folded events shows that four harmonics are sufficient to fully describe the broad thermal pulsations and the folded profiles have the expected statistical properties. Our phase-resolved spectroscopic analyses yield results generally consistent with previous results.  

Overall, we find that the \textit{NICER} data of the four MSPs do not exhibit any anomalies and are thus suitable for detailed inference analyses. The \textit{NICER} data for PSR~J0030$+$0451 presented here are used in \cite{miller19} and \cite{riley19} to obtain estimates on  $M$ and $R$ through principled Bayesian inference analyses. In \cite{miller19}, constraints are presented on the dense matter EoS, as well, while in \cite{raaijmakers19}, the $M$-$R$ constraints from \cite{riley19} are used to provide estimates on the properties of cold, dense matter. Similar investigations for the other three MSPs are ongoing and will be presented in subsequent publications. 

\begin{figure}[h]
\begin{center}
\includegraphics[clip,angle=0,width=0.43\textwidth]{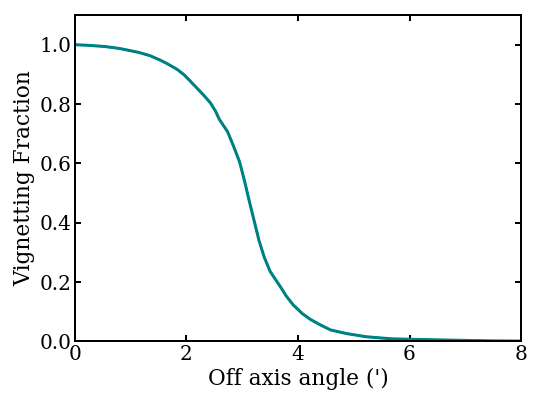}
\caption{\textit{NICER} vignetting curve estimate based on ray tracing showing the fraction of flux received as a function of off-axis angle.}
\label{fig:vignetting}
\end{center}
\end{figure}

\appendix
\section{Background sources near the targets and \nicer{} pointing optimization}
\label{app:optimize}

For each pulsar, we calculate the optimal pointing position based on the method described below. The goal is to maximize the S/N from the pulsar, defined as
\begin{equation}
{\rm S/N} = \frac{C_{\rm PSR} \times t_{\rm exp}}{\sqrt{t_{\rm exp}\times\left( C_{\rm PSR} + \sum_{i} C_{i} + C_{\rm BKG} \right)}},
\end{equation}
by minimizing the contributions $C_{i}$ from nearby sources, while keeping the pulsar count rate $C_{\rm PSR}$ as large as possible. $C_{\rm BKG}$ corresponds to the non-astrophysical (particle and instrumental) background assumed to be constant, and $t_{\rm exp}$ is the exposure time.  The vignetting function of \nicer{} remains relatively flat within $\sim 2\arcmin$ of the aimpoint, but drops sharply at distance $\gtrsim 3\arcmin$ from the aimpoint (see Figure~\ref{fig:vignetting}, which shows the pre-launch estimate based on ray tracing). The optimal pointing method therefore naturally attempts keeping the brightest nearby source outside $\sim$2--3\arcmin\ while maintaining the pulsar within $<2\arcmin$ of the aimpoint.

\begin{figure}[h]
\begin{center}
\includegraphics[clip,angle=0,width=0.4\textwidth]{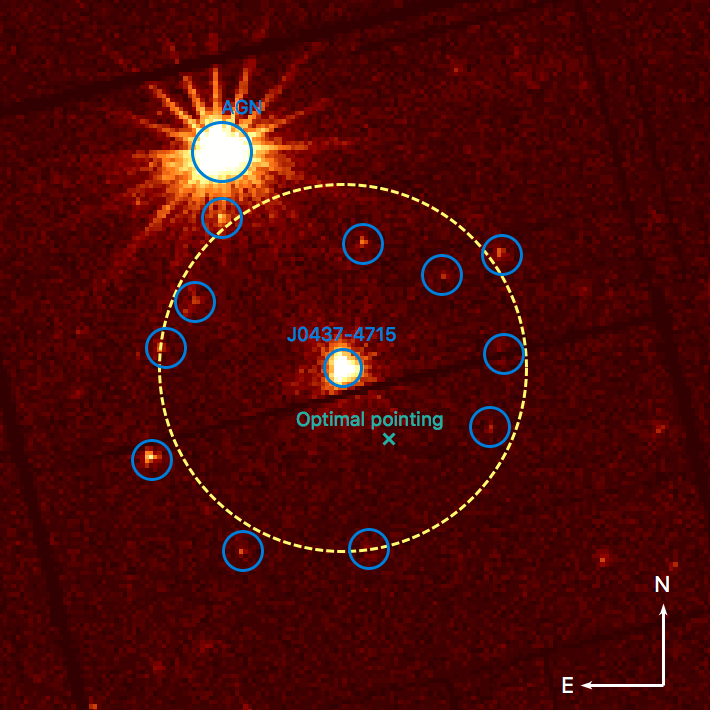}
\includegraphics[clip,angle=0,width=0.4\textwidth]{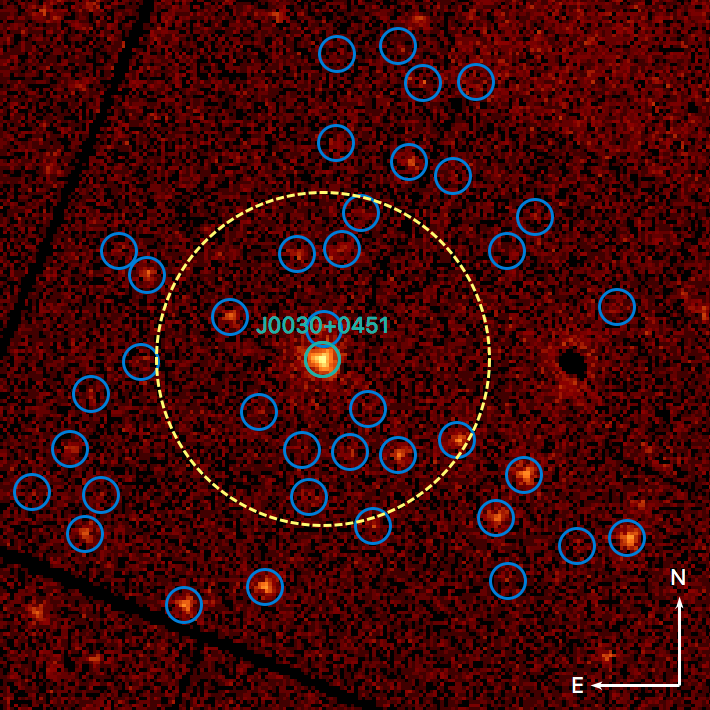}
\includegraphics[clip,angle=0,width=0.4\textwidth]{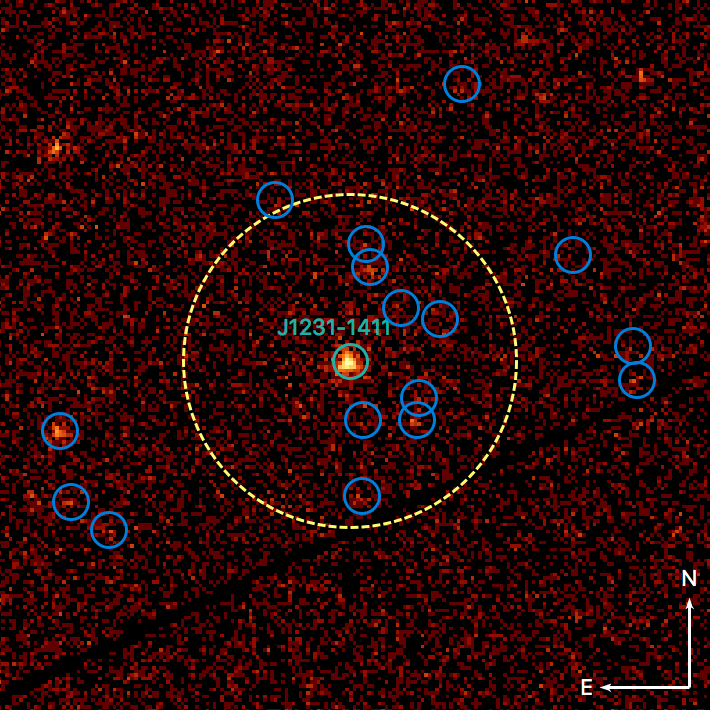}
\includegraphics[clip,angle=0,width=0.4\textwidth]{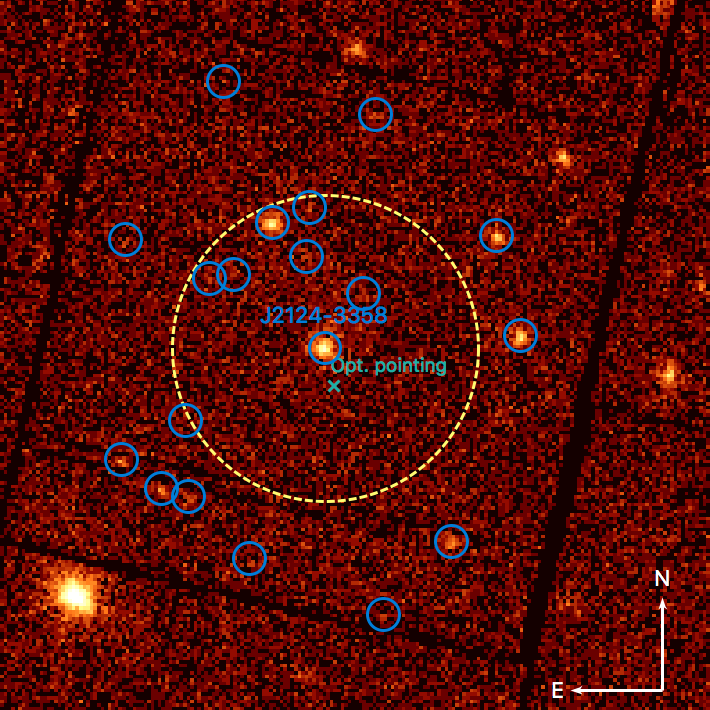}
\caption{\textit{XMM-Newton} EPIC MOS1/2 images of PSRs~J0437$-$4715 (top left), J0030$+$0451 (top right), J1231$-$1411 (bottom left), and J2124$-$3358 (bottom right) and nearby sources marked with blue circles.  The large dashed yellow circle indicates the \nicer{} point spread function with half-power diameter of 6.2\arcmin. For PSRs~J0437$-$4715 and J2124$-$3358, the teal '$\times$' shows the position of the optimal \textit{NICER} pointing that maximizes the S/N from the pulsar, i.e., minimizes the contamination from surrounding sources. For the other two pulsars, PSRs~J0030$+$0451 and J1231$-$1411, the optimal pointing distance from the pulsar and the gain in S/N are negligible.
}
\label{fig:images}
\end{center}
\end{figure}

Since all the pulsars presented in this paper have been observed with the imaging MOS detectors of \textit{XMM-Newton} (Figure~\ref{fig:images}), fluxes and spectral information in the soft X-ray band exist for all nearby sources. We make use of the 3XMM-DR8 Catalog \citep{2016A&A...590A...1R} to estimate the nominal \nicer{} count rate of nearby sources (i.e., if they were placed at the aimpoint). Catalogued sources with $F_{\rm X}\gtrsim 1\tee{-14}$ erg\,s$^{-1}$\,cm$^{-2}$ can be fit with the online spectral fitting tools of the 3XMM-DR8 Catalog\footnote{Available at \url{http://xmm-catalog.irap.omp.eu}.}. For these, we fit the spectrum with simple models (absorbed power-law or thermal models), and use the best fit parameters in WebPIMMS to estimate their \nicer{} count rates. For the fainter sources, when the online spectral fitting tool is not available, we simply convert the 0.2--10~keV reported in the catalog to a \nicer{} count rate assuming a $\Gamma=2$ power-law model.

\begin{figure}[h]
\begin{center}
\includegraphics[clip,angle=0,scale=0.32]{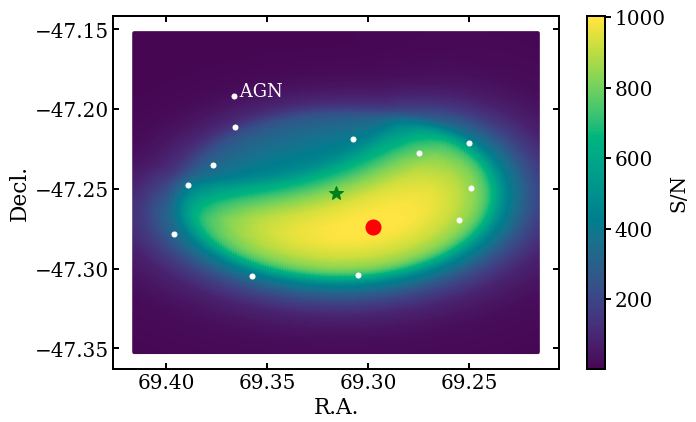}~~~~~
\includegraphics[clip,angle=0,scale=0.32]{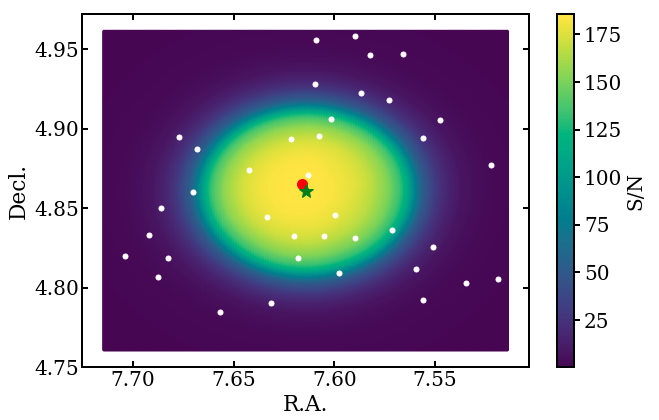}
\includegraphics[clip,angle=0,scale=0.32]{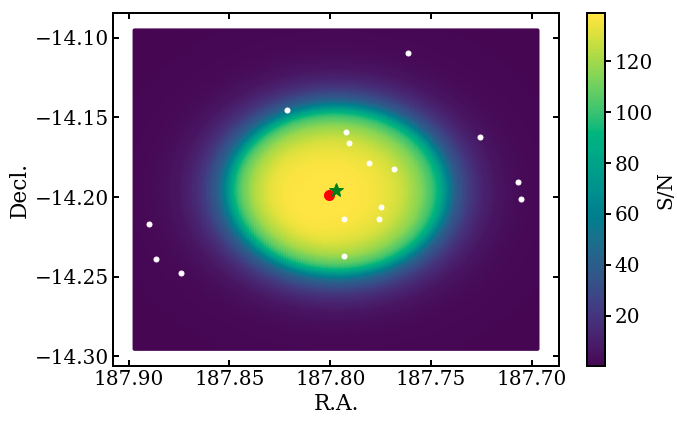}~~~~~
\includegraphics[clip,angle=0,scale=0.32]{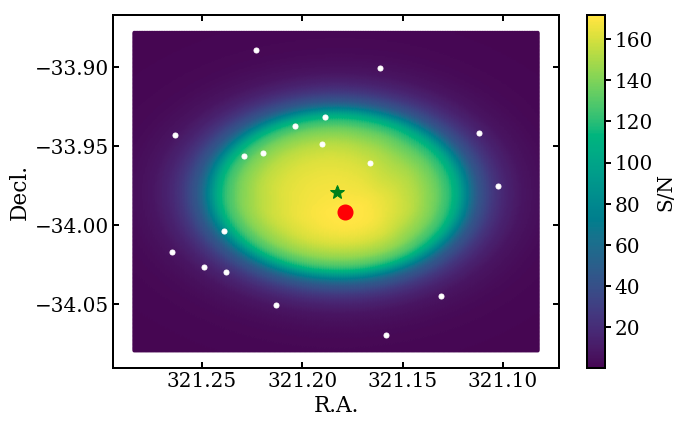}
\caption{\small{Maps of the S/N of PSRs~J0437$-$4715 (top left), J0030+0451 (top right), J1231$-$1411 (bottom left), and J2124$-$3358 (bottom right) as a function of the \textit{NICER} pointing.  The green star indicates the pulsar position, and the red circle shows the calculated optimal pointing position that maximizes the signal to noise ratio. For PSRs~J0437$-$4715, the pointing offset is 1\farcm5 from the pulsar to minimize contamination by the active galactic nucleus RX J0437.4$-$4711, and for PSR~J2124$-$3358 the pointing offset is 1\arcmin.}}
\label{fig:pointings}
\end{center}
\end{figure}

Using the vignetting function for \nicer{}, the actual observed count rate of each source is calculated given its distance from the aim point. With these, we can perform a grid search of alternative pointing around the pulsar to search for that which will maximize the S/N for the pulsar, i.e., which will minimize the contamination from nearby sources without decreasing too much the count rate of the pulsar\footnote{The code is available here: \url{https://github.com/sguillot/NICER}}.  Figure~\ref{fig:pointings} shows the maps of S/N for different pointing around the pulsars, as well as the positions of nearby sources, and the position of the optimal pointing resulting from the grid search described above.

For PSR~J0437$-$4715, we only include sources within 6\arcmin\ and $F>1\times 10^{-14}$ erg\,s$^{-1}$\,cm$^{-2}$; fainter sources will have a negligible contributions for the purpose of optimizing the pointing since the pulsar flux is $F\approx 1\times 10^{-12}$ erg\,s$^{-1}$\,cm$^{-2}$.  Table~\ref{tab:J0437xmm} summarizes the properties and count rates of these nearby sources. The AGN RX~J0437.4$-$4711 dominates the S/N map, and the optimal pointing is naturally in the direction opposite to the AGN position.  For all other pulsars, we considered all sources within 6\arcmin\ of the targets. Table~\ref{tab:J0030xmm}--\ref{tab:J2124xmm} present the properties of the nearby sources, and the S/N maps is displayed in Figures~\ref{fig:pointings}.

\begin{deluxetable}{lcccccc}
\tabletypesize{\footnotesize}
\tablecolumns{8} 
\tablewidth{0pt}  
\tablecaption{Other X-ray sources near PSR~J0437--4715  \label{tab:J0437xmm}} 
\tablehead{
\colhead{Name} & \colhead{$F_{\rm 0.2-10\,keV}$} & \colhead{\nicer} & \colhead{Distance} & \colhead{Vignetting} & \colhead{Scaled \nicer} & \colhead{Count rate at} \\
\colhead{}  &\colhead{($\times10^{-14}$} & \colhead{count rate} & \colhead{from PSR} & \colhead{fraction} & \colhead{count rate} & \colhead{optimal pointing} \\
\colhead{ } & \colhead{ erg\,s$^{-1}$\,cm$^{-2}$)} & \colhead{(s$^{-1}$)} & \colhead{(arcmin)} & \colhead{} & \colhead{(s$^{-1}$)} & \colhead{(s$^{-1}$)}
} 
\startdata
3XMM\,J043728.1$-$471129 (AGN) & 1300& 9.577 & 4.18 & 0.07 & 0.7629 & 0.0768 \\
3XMM\,J043735.1$-$471638       & 6.4 & 0.037 & 3.62 & 0.21 & 0.0075 & 0.0038 \\
3XMM\,J043728.0$-$471237       & 3.1 & 0.021 & 3.20 & 0.42 & 0.0088 & 0.0007 \\
3XMM\,J043734.1$-$471448       & 3.2 & 0.007 & 2.99 & 0.50 & 0.0040 & 0.0007 \\
3XMM\,J043730.6$-$471400       & 2.9 & 0.019 & 2.70 & 0.68 & 0.0136 & 0.0020 \\
3XMM\,J043705.9$-$471336       & 2.5 & 0.003 & 2.25 & 0.83 & 0.0025 & 0.0018 \\
3XMM\,J043700.0$-$471454       & 2.5 & 0.009 & 2.73 & 0.71 & 0.0064 & 0.0071 \\
3XMM\,J043701.3$-$471609       & 2.3 & 0.002 & 2.69 & 0.72 & 0.0014 & 0.0019 \\
3XMM\,J043713.3$-$471812       & 1.6 & 0.006 & 3.14 & 0.50 & 0.0028 & 0.0055 \\
3XMM\,J043700.4$-$471313       & 1.4 & 0.011 & 3.27 & 0.34 & 0.0041 & 0.0020 \\
3XMM\,J043725.9$-$471814       & 1.2 & 0.006 & 3.56 & 0.24 & 0.0013 & 0.0032 \\
3XMM\,J043714.0$-$471301       & 1.1 & 0.013 & 2.06 & 0.87 & 0.0114 & 0.0040 \\
\enddata
\end{deluxetable}

\begin{deluxetable}{lcccccc}
\tabletypesize{\footnotesize}
\tablecolumns{7} 
\tablewidth{0pt}  
\tablecaption{Other X-ray sources near PSR~J0030+0451
\label{tab:J0030xmm}} 
\tablehead{
\colhead{Name} & \colhead{$F_{\rm 0.2-10\,keV}$} & \colhead{\nicer} & \colhead{Distance} & \colhead{Vignetting} & \colhead{Scaled \nicer} & \colhead{Count rate at} \\
\colhead{}  &\colhead{($\times10^{-14}$} & \colhead{count rate} & \colhead{from PSR} & \colhead{fraction} & \colhead{count rate} & \colhead{optimal pointing} \\
\colhead{ } & \colhead{ erg\,s$^{-1}$\,cm$^{-2}$)} & \colhead{(s$^{-1}$)} & \colhead{(arcmin)} & \colhead{} & \colhead{(s$^{-1}$)} & \colhead{(s$^{-1}$)}
}
\startdata 
3XMM\,J003027.2+045214 & 1.00 & 0.004 & 0.59 & 0.99 & 0.004   & 0.004   \\
3XMM\,J003023.9+045044 & 0.43 & 0.002 & 1.27 & 0.96 & 0.002   & 0.002   \\ 
3XMM\,J003032.0+045041 & 0.27 & 0.001 & 1.51 & 0.93 & 0.001   & 0.001   \\
3XMM\,J003028.8+044958 & 0.91 & 0.004 & 1.73 & 0.91 & 0.003   & 0.004   \\
3XMM\,J003025.2+044956 & 1.04 & 0.004 & 1.80 & 0.91 & 0.004   & 0.003   \\
3XMM\,J003034.2+045227 & 1.22 & 0.005 & 1.87 & 0.90 & 0.006   & 0.005   \\
3XMM\,J003029.2+045337 & 1.37 & 0.006 & 2.01 & 0.89 & 0.005   & 0.006   \\
3XMM\,J003025.8+045343 & 0.71 & 0.003 & 2.10 & 0.86 & 0.003   & 0.003   \\
3XMM\,J003021.6+044953 & 2.38 & 0.010 & 2.28 & 0.80 & 0.008   & 0.008   \\
3XMM\,J003028.3+044906 & 0.79 & 0.003 & 2.56 & 0.71 & 0.002   & 0.002   \\
3XMM\,J003024.4+045423 & 0.74 & 0.003 & 2.83 & 0.62 & 0.002   & 0.002   \\
3XMM\,J003017.2+045010 & 3.17 & 0.013 & 2.95 & 0.58 & 0.008   & 0.006   \\
3XMM\,J003023.5+044834 & 0.47 & 0.002 & 3.24 & 0.45 & 0.001   & 0.0005  \\
3XMM\,J003040.8+045137 & 0.76 & 0.003 & 3.35 & 0.40 & 0.001   & 0.001   \\
3XMM\,J003040.4+045314 & 3.81 & 0.016 & 3.62 & 0.28 & 0.004   & 0.005   \\
3XMM\,J003013.5+045340 & 2.17 & 0.009 & 4.00 & 0.11 & 0.001   & 0.001   \\
3XMM\,J003020.8+045520 & 1.50 & 0.006 & 4.03 & 0.11 & 0.001   & 0.0008  \\
3XMM\,J003026.3+045541 & 0.56 & 0.002 & 4.03 & 0.11 & 0.0002  & 0.0003  \\
3XMM\,J003017.5+045504 & 1.08 & 0.004 & 4.21 & 0.09 & 0.0004  & 0.0004  \\
3XMM\,J003042.5+045341 & 0.44 & 0.002 & 4.28 & 0.08 & 0.0002  & 0.0002  \\
3XMM\,J003012.2+044931 & 5.83 & 0.024 & 4.35 & 0.08 & 0.002   & 0.0009  \\
3XMM\,J003044.6+045101 & 0.23 & 0.001 & 4.35 & 0.08 & 0.0001  & 0.00007 \\
3XMM\,J003031.6+044726 & 4.21 & 0.017 & 4.35 & 0.08 & 0.001   & 0.0007  \\
3XMM\,J003014.3+044843 & 2.84 & 0.012 & 4.39 & 0.07 & 0.001   & 0.0004  \\
3XMM\,J003011.4+045419 & 0.28 & 0.001 & 4.79 & 0.04 & 0.0001  & 0.00003 \\
3XMM\,J003043.8+044908 & 0.34 & 0.001 & 4.82 & 0.04 & 0.0001  & 0.00003 \\
3XMM\,J003046.1+044959 & 0.92 & 0.004 & 4.95 & 0.02 & 0.0001  & 0.00009 \\
3XMM\,J003037.6+044705 & 3.34 & 0.014 & 5.23 & 0.02 & 0.0003  & 0.0002  \\
3XMM\,J003013.4+044732 & 0.43 & 0.002 & 5.39 & 0.02 & 0.00003 & 0.00002 \\
3XMM\,J003045.0+044825 & 2.95 & 0.012 & 5.46 & 0.02 & 0.0002  & 0.0001  \\
3XMM\,J003019.8+045648 & 0.19 & 0.001 & 5.49 & 0.02 & 0.00001 & 0.00001 \\
3XMM\,J003005.3+045238 & 0.26 & 0.001 & 5.57 & 0.01 & 0.00002 & 0.00001 \\
3XMM\,J003026.2+045721 & 0.74 & 0.003 & 5.70 & 0.01 & 0.00004 & 0.00003 \\
3XMM\,J003008.3+044811 & 0.30 & 0.001 & 5.88 & 0.01 & 0.00001 & 0.00001 \\
3XMM\,J003015.8+045649 & 2.66 & 0.011 & 5.91 & 0.01 & 0.0001  & 0.00008 \\ 
3XMM\,J003049.0+044911 & 0.26 & 0.001 & 5.93 & 0.01 & 0.00001 & 0.00001 \\ 
3XMM\,J003021.6+045729 & 1.26 & 0.005 & 6.00 & 0.01 & 0.00003 & 0.00004 \\
3XMM\,J003004.5+044820 & 8.13 & 0.034 & 6.61 & 0.01 & 0.0001  & 0.0001  \\
\enddata
\end{deluxetable}

\begin{deluxetable}{lcccccc}
\tablecolumns{7}
\tabletypesize{\footnotesize}
\tablecaption{Other X-ray sources near PSR~J1231--1411
\label{tab:J1231xmm}} 
\tablehead{
\colhead{Name} & \colhead{$F_{\rm 0.2-10\,keV}$} & \colhead{\nicer} & \colhead{Distance} & \colhead{Vignetting} & \colhead{Scaled \nicer} & \colhead{Count rate at} \\
\colhead{}  &\colhead{($\times10^{-14}$} & \colhead{count rate} & \colhead{from PSR} & \colhead{fraction} & \colhead{count rate} & \colhead{optimal pointing} \\
\colhead{ } & \colhead{ erg\,s$^{-1}$\,cm$^{-2}$)} & \colhead{(s$^{-1}$)} & \colhead{(arcmin)} & \colhead{} & \colhead{(s$^{-1}$)} & \colhead{(s$^{-1}$)}
}
\startdata
3XMM\,J123110.3$-$141249 & 0.55 & 0.002 & 1.12 & 0.97 & 0.0022 & 0.0020 \\ 
3XMM\,J123107.4$-$141044 & 0.76 & 0.003 & 1.36 & 0.95 & 0.0030 & 0.0028 \\ 
3XMM\,J123106.0$-$141224 & 0.50 & 0.002 & 1.45 & 0.94 & 0.0019 & 0.0019 \\ 
3XMM\,J123106.2$-$141249 & 2.68 & 0.011 & 1.65 & 0.92 & 0.0104 & 0.0102 \\ 
3XMM\,J123109.8$-$140958 & 2.67 & 0.011 & 1.78 & 0.91 & 0.0102 & 0.0097 \\ 
3XMM\,J123104.4$-$141056 & 5.00 & 0.021 & 1.85 & 0.90 & 0.0190 & 0.0180 \\ 
3XMM\,J123110.1$-$140933 & 0.38 & 0.002 & 2.19 & 0.83 & 0.0013 & 0.0016 \\ 
3XMM\,J123110.4$-$141414 & 1.02 & 0.004 & 2.52 & 0.72 & 0.0031 & 0.0029 \\ 
3XMM\,J123117.1$-$140843 & 0.68 & 0.003 & 3.32 & 0.42 & 0.0012 & 0.0008 \\ 
3XMM\,J123054.2$-$140945 & 0.74 & 0.003 & 4.57 & 0.06 & 0.0002 & 0.00008\\ 
3XMM\,J123049.6$-$141126 & 1.16 & 0.005 & 5.25 & 0.02 & 0.0001 & 0.00006\\ 
3XMM\,J123049.3$-$141204 & 1.90 & 0.008 & 5.34 & 0.02 & 0.0001 & 0.00008\\ 
3XMM\,J123129.8$-$141452 & 0.79 & 0.003 & 5.50 & 0.02 & 0.0001 & 0.00005\\ 
3XMM\,J123102.7$-$140634 & 1.85 & 0.008 & 5.55 & 0.01 & 0.0001 & 0.00006\\ 
3XMM\,J123133.6$-$141301 & 3.18 & 0.013 & 5.56 & 0.01 & 0.0002 & 0.00016\\ 
3XMM\,J123132.7$-$141420 & 1.34 & 0.006 & 5.81 & 0.01 & 0.0001 & 0.00006\\
\enddata
\end{deluxetable}

\begin{deluxetable}{lcccccc}
\tablecolumns{7} 
\tabletypesize{\footnotesize}
\tablewidth{0pt}  
\tablecaption{Other X-ray sources near PSR~J2124--3358
\label{tab:J2124xmm}} 
\tablehead{
\colhead{Name} & \colhead{$F_{\rm 0.2-10\,keV}$} & \colhead{\nicer} & \colhead{Distance} & \colhead{Vignetting} & \colhead{Scaled \nicer} & \colhead{Count rate at} \\
\colhead{}  &\colhead{($\times10^{-14}$} & \colhead{count rate} & \colhead{from PSR} & \colhead{fraction} & \colhead{count rate} & \colhead{optimal pointing} \\
\colhead{ } & \colhead{ erg\,s$^{-1}$\,cm$^{-2}$)} & \colhead{(s$^{-1}$)} & \colhead{(arcmin)} & \colhead{} & \colhead{(s$^{-1}$)} & \colhead{(s$^{-1}$)}
} 
\startdata
3XMM\,J212426.9$-$335629 & 4.11 & 0.026 & 4.15 & 0.08 & 0.0022 & 0.0010 \\
3XMM\,J212424.6$-$335830 & 5.84 & 0.040 & 4.02 & 0.11 & 0.0042 & 0.0035 \\
3XMM\,J212448.8$-$335613 & 11.2 & 0.033 & 2.76 & 0.71 & 0.0231 & 0.0091 \\
3XMM\,J212431.3$-$340240 & 1.91 & 0.011 & 4.72 & 0.03 & 0.0004 & 0.0009 \\
3XMM\,J212459.6$-$340135 & 1.70 & 0.007 & 4.30 & 0.06 & 0.0005 & 0.0009 \\
3XMM\,J212503.5$-$340100 & 1.42 & 0.009 & 4.66 & 0.03 & 0.0003 & 0.0005 \\   
3XMM\,J212438.7$-$335401 & 0.97 & 0.005 & 4.84 & 0.03 & 0.0001 & 0.0001 \\
3XMM\,J212457.0$-$340145 & 1.52 & 0.008 & 4.07 & 0.09 & 0.0008 & 0.0014 \\
3XMM\,J212439.9$-$335739 & 1.27 & 0.007 & 1.37 & 0.96 & 0.0067 & 0.0063 \\
3XMM\,J212452.6$-$335716 & 0.82 & 0.004 & 2.35 & 0.82 & 0.0037 & 0.0025 \\
3XMM\,J212437.9$-$340409 & 1.37 & 0.008 & 5.53 & 0.01 & 0.0001 & 0.0002 \\
3XMM\,J212445.5$-$335654 & 0.25 & 0.001 & 1.88 & 0.91 & 0.0013 & 0.0010 \\
3XMM\,J212445.2$-$335554 & 0.42 & 0.002 & 2.86 & 0.65 & 0.0015 & 0.0005 \\
3XMM\,J212454.9$-$335721 & 1.27 & 0.007 & 2.69 & 0.72 & 0.0050 & 0.0024 \\
3XMM\,J212503.1$-$335633 & 0.36 & 0.002 & 4.57 & 0.04 & 0.0001 & 0.0000 \\
3XMM\,J212453.5$-$335321 & 0.89 & 0.005 & 5.75 & 0.01 & 0.0000 & 0.0000 \\
3XMM\,J212457.3$-$340013 & 0.52 & 0.003 & 3.15 & 0.45 & 0.0013 & 0.0014 \\
3XMM\,J212451.0$-$340301 & 0.69 & 0.004 & 4.53 & 0.04 & 0.0002 & 0.0005 \\
\enddata
\end{deluxetable}

\acknowledgements

This work was supported by NASA through the \textit{NICER} mission and the Astrophysics Explorers Program. A portion of the analysis presented was based on archival observations obtained with \textit{XMM-Newton}, an ESA science mission with instruments and contributions directly funded by ESA Member States and NASA. S.G. acknowledges the support of the Centre National d'\'{E}tudes Spatiales (CNES).  A.L.W. and T.E.R. acknowledge support from ERC Starting Grant No.~639217 CSINEUTRONSTAR (PI Watts). M.C.M. is grateful for the hospitality of Perimeter Institute where part of
this work was carried out. Research at Perimeter Institute is supported in part by the Government of Canada through the Department of Innovation, Science and Economic Development Canada and by the Province of Ontario through the Ministry of Economic Development, Job Creation and Trade. This research has made use of data and software provided by the High Energy Astrophysics Science Archive Research Center (HEASARC), which is a service of the Astrophysics Science Division at NASA/GSFC and the High Energy Astrophysics Division of the Smithsonian  Astrophysical Observatory.  We acknowledge extensive use of NASA's Astrophysics Data System (ADS) Bibliographic Services and the ArXiv. 

\software{
    \texttt{HEAsoft} \citep{2014ascl.soft08004N}, \texttt{Tempo2} \citep{2006MNRAS.369..655H}, \texttt{PINT} (\url{https://github.com/nanograv/pint}), \texttt{XSPEC} \citep{arnaud96}), \texttt{NICERSoft} (\url{https://github.com/paulray/NICERsoft}).
    }

\facilities{\textit{NICER}, \textit{XMM-Newton}}

\bibliographystyle{aasjournal}
\bibliography{wp_references}


\end{document}